# Modified deformation behaviour of self-ion irradiated tungsten: A combined nano-indentation, HR-EBSD and crystal plasticity study

Suchandrima Das[a*], Hongbing Yu[a], Kenichiro Mizohata[b], Edmund Tarleton[a,c], Felix Hofmann[a†]

[a]*Department of Engineering Science, University of Oxford, Parks Road, Oxford OX1 3PJ, UK*
[b]*Accelerator Laboratory, University of Helsinki, P.O. Box 64, 00560 Helsinki, Finland*
[b]*Department of Materials, University of Oxford, Parks Road, Oxford OX1 3PH, UK*

[*]*suchandrima.das@eng.ox.ac.uk*

[†]*felix.hofmann@eng.ox.ac.uk*

## Keywords

B. crystal plasticity; B. polycrystalline material; C. electron microscopy; C. non-destructive evaluation; C. mechanical testing; irradiation damage; ion implantation; nano-indentation

## Abstract

Predicting the dramatic changes in mechanical and physical properties caused by irradiation damage is key for the design of future nuclear fission and fusion reactors. Self-ion irradiation provides an attractive tool for mimicking the effects of neutron irradiation. However, the damaged layer of self-ion implanted samples is only a few microns thick, making it difficult to estimate macroscopic properties. Here we address this challenge using a combination of experimental and modelling techniques. We concentrate on self-ion-implanted tungsten, the front-runner for fusion reactor armour components and a prototypical bcc material. To capture dose-dependent evolution of properties, we experimentally characterise samples with damage levels from 0.01 to 1 dpa. Spherical nano-indentation of <001> grains shows hardness increasing up to a dose of 0.032 dpa, beyond which it saturates. Atomic force microscopy (AFM) measurements show pile-up increasing up to the same dose, beyond which large pile-up and slip-steps are seen. Based on these observations we develop a simple crystal plasticity finite element (CPFE) model for the irradiated material. It captures irradiation-induced hardening followed by strain-softening through the interaction of irradiation-induced-defects and gliding dislocations. The shear resistance of irradiation-induced-defects is physically-based, estimated from transmission electron microscopy (TEM) observations of similarly irradiated samples. Nano-indentation of pristine tungsten and implanted tungsten of doses 0.01, 0.1, 0.32 and 1 dpa is simulated. Only two model parameters are fitted to the experimental results of the 0.01 dpa sample and are kept unchanged for all other doses. The peak indentation load, indent surface profiles and damage saturation predicted by the CPFE model closely match our experimental observations.

Predicted lattice distortions and dislocation distributions around indents agree well with corresponding measurements from high-resolution electron backscatter diffraction (HR-EBSD). Finally, the CPFE model is used to predict the macroscopic stress-strain response of similarly irradiated bulk tungsten material. This macroscopic information is the key input required for design of fusion armour components.

## 1. Introduction

The lifetime of plasma-facing armour components in future fusion reactors will be compromised due to in-service irradiation by fusion neutrons. (Entler et al., 2018; Knaster et al., 2016). Gaseous elements such as hydrogen and helium will also be incorporated into the armour components either directly from the fusion plasma or formed as a result of neutron-irradiation-induced transmutation (Gilbert et al., 2012; Hammond, 2017). Irradiation causes significant changes in the physical and mechanical properties, such as increased hardening, embrittlement, dimensional change, residual stress, reduced thermal conductivity, etc. (Armstrong et al., 2013; S. Das et al., 2018a, 2018b; Fang et al., 2018; Hofmann et al., 2015a; Reza et al., 2020a; Yi et al., 2013). Making armour components more resistant to radiation and accurate prediction of their lifetime will be essential for realising commercial fusion power. This requires fundamental study of the change-inducing factors; nucleation and evolution of irradiation-induced defects, and their interaction for example with glide dislocations. Since fusion reactor conditions cannot be recreated yet, such studies are commonly done using representative models of inactive, ion-implanted metals. For example, helium-implanted and self-ion-implanted materials are used as representative models of helium-irradiated and neutron-irradiated materials respectively (Armstrong et al., 2011; DeBroglie et al., 2015; Gibson et al., 2015; Heintze et al., 2011).

Ion-implantation is inexpensive and allows isolated investigation of displacement damage effects up to high dose levels (a few hundred displacements per atom (dpa)), without the added complexity of transmutation and gas evolution (Dennett et al., 2018; Toloczko et al., 2014). However, the ion-implanted damage layer is only a few microns thick (precise depth depends on the ion energy used and the atomic number of the ions and target material; two isotopes with the same atomic number would yield distinct damage). Thus, the ion-implanted layer is too thin for conventional, macroscopic testing approaches. Nano-indentation is commonly used to probe the mechanical properties of these thin ion-implanted layers (Heintze et al., 2009; Hosemann et al., 2009; N. Li et al., 2009; Oliver and Pharr, 2004). Further characterisation of the deformation behaviour of the ion-implanted layer, can be done by examining the deformation field around and beneath indents using techniques such as high-resolution electron back-scattered diffraction (HR-EBSD) ((Wilkinson, 1996), 3D Laue diffraction (S. Das et al., 2018a), high-resolution digital image correlation (HR-DIC) (Guan et al., 2017), etc. Additionally examination by transmission electron microscopy (TEM) can give direct information about the defect microstructure induced by implantation (Yi, 2013). Such experimental characterisation can be performed on ion-implanted samples with systematically varying crystallographic orientation, varying damage dose, varying temperature conditions or different impurity concentrations. Through these studies the influence of these different parameters on the formation and structure of irradiation-induced defects and resulting changes in material properties can be probed.

Information gathered from the combination of these characterisation techniques corresponds to the thin irradiated layer. However, for reactor design, it is essential to "translate" this information to estimate the deformation behaviour of a similarly irradiated macroscopic polycrystal. Previously, several efforts have been made in this regard: Extraction of stress-strain curves from raw data acquired through spherical nano-indentation measurements have been found particularly useful (Pathak and Kalidindi, 2015). For example, Pathak et al. used this technique to analyse local loading and unloading elastic moduli, the local indentation yield strengths, and the post-yield strain hardening behaviour of pure and helium-implanted tungsten (Pathak et al., 2017). Further, these authors demonstrated that by varying the indenter size, the heterogeneous characteristics of the radiation-induced damage zone can be probed and directly correlated with the local material structure obtained from EBSD and/or TEM. Nano-indentation and micro-compression testing, combined with the Nix and Gao model or the Dao model, has also been found useful to estimate the yield strength of irradiated materials (Hosemann et al., 2008). The reliability of the scaled data from nano-indentation was assessed by Krumwiede et al. by comparing macroscopic tensile test data (yield and flow stresses) to data acquired from nano-indentation, for different neutron-irradiated materials (Krumwiede et al., 2018). It was seen that on average the neutron-irradiated condition tensile strength could be predicted within ~15% and the flow stress within ~5%. However, the study also indicates that application of similar techniques for ion-irradiated materials would involve much larger uncertainties. Similar scaling techniques have also been applied to micro-pillar compression or micro-tension measurements to estimate macroscopic properties of irradiated materials. While such direct scaling of nano-indentation or uniaxial tests is cost- and time-effective, there remain some associated challenges. For example, inconsistencies have been found between the stress-strain curves obtained from nano-indentation and those found from uniaxial (such as micropillar or micro-compression) tests for proton-irradiated stainless-steel (Weaver et al., 2017).

Here, we propose an alternative method of translating the micro-mechanical experimental data to predict bulk behaviour, by using the small-scale response information from the thin ion-implanted layer to develop a mesoscale model of the ion-damaged material. Importantly this must capture the key physics controlling property change, accounting for example for the interaction between expected defect-types and gliding dislocations, considering orientation-dependence, etc. Furthermore, the model parameters should be linked to dose-related observables, such as defect density obtained from TEM. Once tuned and validated against experimental observations, this mesoscale model can then be used to predict the deformation response of a similarly irradiated macroscopic polycrystal; a task challenging to achieve experimentally owing to the limited ion-penetration depth. Here we demonstrate all the parts of this process using tungsten as a prototypical material. The predicted macroscopic behaviour of ion-irradiated materials could then be used to inform the design of fusion devices. Of course, differences between actual fusion conditions and ion-irradiation must be accounted for in this context, e.g. in terms of primary knock-on atom (PKA) recoil spectrum, transmutation, displacement rates and the effects of environmental parameters such as stress-state, flux pulsing and temperature.

Tungsten is the most promising material for plasma-facing armour components in future fusion reactors (Maisonnier et al., 2005; Suchandrima Das, 2019; Wei et al., 2014). Its high melting point (3422 °C), low tritium-retention rate, low sputtering rate and good thermal conductivity, make it suitable for withstanding the harsh conditions anticipated in service. However, past studies show undesirable irradiation-induced changes in its properties. In-situ TEM of tungsten implanted with 150 keV self-ions, at room temperature, showed that the first observable defects (predominantly ½<111> vacancy loops) appear at very low doses < 0.01 dpa. Their concentration increases almost linearly with dose before saturating at higher doses (0.1 - 1 dpa) (Yi et al., 2016). Nano-indentation of tungsten-implanted tungsten layers have shown that these defects can cause significant hardening (Armstrong et al., 2013; Gibson et al., 2014), with suppression of pile-up around indents indicating a substantial strain hardening (Armstrong et al., 2011). Also, reduction in thermal diffusivity and ductility is expected to be caused by irradiation defects (Hofmann et al., 2015a; Reza et al., 2020b; Zinkle and Was, 2013).

To integrate these changes into a material model, quantitative understanding of the underlying mechanisms is required. However, such understanding is difficult to derive conclusively from past studies: For example, the suppression in pile-up noticed around indents in self-ion-implanted tungsten is unexplained (Armstrong et al., 2011). This is particularly surprising, as it is in stark contrast to indents in helium-implanted tungsten which show a large increase in pileup (alongside irradiation hardening), and slip channels indicative of strain-softening (Beck et al., 2017; S. Das et al., 2018a). This raises the question as to whether the interaction of gliding dislocations with implantation defects in tungsten-implanted tungsten is different from that in the helium-implanted case and why this would be so.

Here we aim to address these questions and develop an understanding of the physics of irradiation induced changes, by characterising self-ion-implanted tungsten samples using scanning electron microscopy (SEM), atomic force microscopy (AFM), HR-EBSD and TEM data. We consider polycrystalline tungsten samples (99.99% purity), implanted at room temperature and exposed to a range of damage levels (0.01, 0.1, 0.32 and 1 dpa). Based on our understanding from experimental results, we develop a material model representative of the self-ion implanted tungsten at varying doses. We verify the material model formulation by comparing its predictions of deformation behaviour with corresponding experimental results. Finally, we use the verified material model to predict the deformation behaviour of similarly irradiated macroscopic self-ion implanted polycrystalline tungsten.

The outlook is to build material models accounting for all possible parameters (like temperature, purity, orientation, dose) influencing the implantation-induced changes in material properties. However, the effect of each parameter must first be considered individually, before incorporating into one model. Considering the complexity of the task, we start here by accounting for two parameters; crystallographic orientation and implantation dose. To avoid influences of other parameters like temperature and impurity, high purity samples implanted at room temperature are considered.

## 2. Experimental Methods
### 2.1. Sample preparation
Seven samples, of size 1 cm square with a thickness of 1 mm were cut from a polycrystalline tungsten sheet (99.99 wt% purity, procured from Plansee) and then annealed at 1500 °C for 24 h in vacuum (~$10^{-5}$ mbar). A high quality, damage free surface finish was obtained by mechanical grinding, polishing with diamond paste and colloidal silica, and finally electropolishing in an electrolyte of 1% NaOH aqueous solution (8 V, room temperature). The electropolished samples had an average grain size of 75 μm.

### 2.2. Ion implantation
High energy ion implantation was performed to mimic neutron-induced damage. Implantation with 20 MeV tungsten ions was used (+5 charge state, 5 MV tandem accelerator (Tikkanen et al., 2004)), to create a relatively homogeneous damage profile across a ~2.5 μm thick layer (Hosemann et al., 2012).

Implantations were carried out at room temperature. The ion beam was raster scanned across the sample area (~15×15 mm$^2$) with a sweeping frequency of ~5 - 10 Hz in both X and Y directions and was slightly defocused to a spot diameter of ~5 mm. Beam current and dose before the target chamber were monitored using a beam profilometer (BPM). The BPM current measurement was calibrated using a faraday cup (f-cup) in the target chamber. A 12.5 mm diameter collimator was placed in front of the f-cup to define the area of the f-cup. The beam current and exposure time were adjusted to obtain the desired damage levels (Table 1). The same flux density was used for all implantations.

Figure 1 (a) shows distribution of damage and ion-ranges estimated using the SRIM code (Ziegler and Biersack, 2010) (quick Kinchin-Pease method, 68 eV displacement energy (ASTM International, West Conshohocken, PA, 2009, 2009)). The damage level listed in Table 1 refers to the peak of the damage profile.

*Table 1 – List of the different damage levels considered and the corresponding implantation fluence and flux used for the self-ion implanted tungsten samples.*

| Dose level (dpa) | Fluence (ions/cm$^2$) | Flux (ions/cm$^2$/s) |
|---|---|---|
| 0.01 | $2.55 \times 10^{12}$ | $3.1 – 5.0 \times 10^{10}$ |
| 0.018 | $4.61 \times 10^{12}$ | $3.1 – 5.0 \times 10^{10}$ |
| 0.032 | $8.20 \times 10^{12}$ | $3.1 – 5.0 \times 10^{10}$ |
| 0.1 | $2.54 \times 10^{13}$ | $3.1 – 5.0 \times 10^{10}$ |
| 0.32 | $8.11 \times 10^{13}$ | $3.1 – 5.0 \times 10^{10}$ |
| 1 | $2.53 \times 10^{14}$ | $3.1 – 5.0 \times 10^{10}$ |

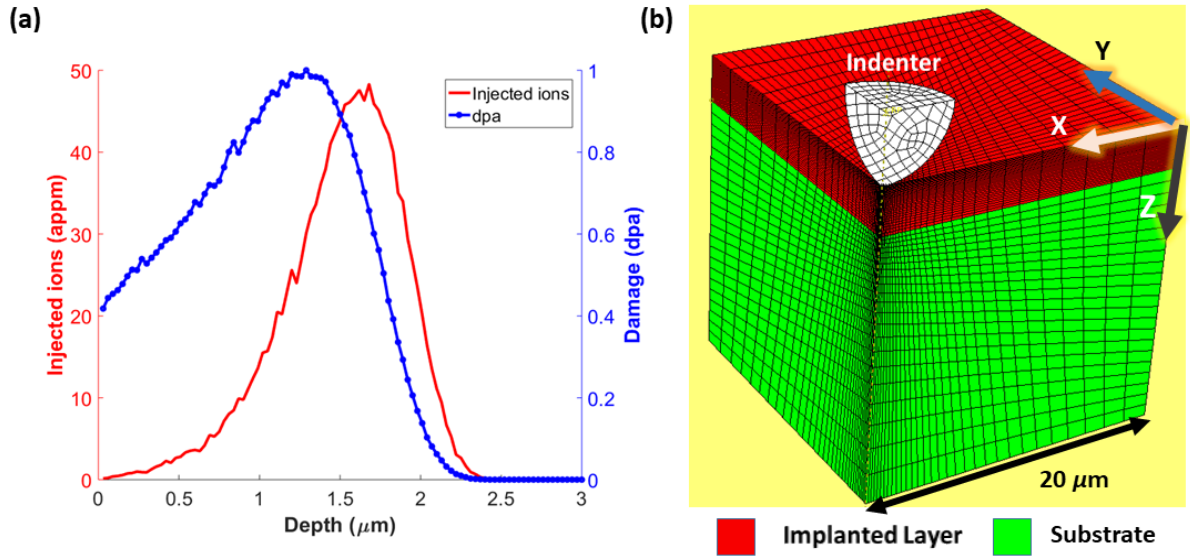

*Figure 1 – (a) SRIM estimate of the injected ion concentration (blue curve) and the implantation induced displacement damage (pink curve) as a function of depth in the tungsten sample damaged to 1 dpa. The profiles for the other damage levels simply correspond to scaled versions of these profiles. (b) CPFE mesh of the quarter model created in Abaqus for simulation of the nano-indentation experiments for the <001> and <011> oriented grains.*

### 2.3. Choice of test points

Electron back scatter diffraction (EBSD) was used to identify three suitable grains with near <001> surface normal orientation in each sample. In addition, a <011> and a <111> oriented grain were also identified in the unimplanted and the 1 dpa sample to investigate orientation dependence of the indentation response. The exact orientations of the chosen grains are listed in Appendix A.

### 2.4. Nano-indentation & Atomic force microscopy

In each of the chosen grains (Table A.1), a 500 nm deep indent was made with a spherical indenter tip of radius ~ 5 µm (MTS NanoXp, Synton diamond tip). Grains were chosen to allow a spacing of at least 50 µm between indents. Between depths 500 nm and 2 µm in the implanted layer, there is significant variation in the injected ion concentration. However, our previous thermal diffusivity measurements on these samples showed that the effect of the injected ions is small compared to the cascade damage they cause (Reza et al., 2020a). Thus, our analysis here concentrates on the cascade damage. Figure 1(a) shows that the corresponding variation in induced damage or displacement per atom (dpa) in the implanted layer is small; specifically, between 0.5 µm and 2 µm, the dpa varies between 0.63 and 1 dpa with average dpa of 0.73 dpa and standard deviation of 0.23 dpa (~20%). Our previous thermal transport measurements suggest that the differences in damage microstructure that arise due to this variation in dose are small (Reza et al., 2020a). As such, we assume the nano-indentation measurements to be unaffected by the depth-dependent variation in the morphology and number density of loops in the implanted layer.

The general consensus is that, to exclude the effects of the underlying substrate, the indentation depth should be limited to 10-20% of the implanted layer thickness (Bhattacharya and Nix, 1988; Heintze et al., 2009; Sawa et al., 1999). Based on the ~2.5 µm thickness of the implanted layer in our samples, the indentation depth should thus be between 250 and 500 nm. Another challenge associated with indentation measurements is the indentation size effect (ISE) (Fleck and Hutchinson, 1997), which becomes more pronounced for smaller indentation depths. Thus, to reduce ISE and to probe the relatively flat damage (dpa) profile beyond 500 nm (Figure 1(a) - as noted above the standard deviation of dpa beyond 500 nm is ~20%), a maximum indentation depth of 500 nm was chosen.

Atomic force microscopy was used to measure the surface height profile in the vicinity of each indent. These measurements were done in contact mode using a Digital Instruments Dimension 3100 AFM with Bruker CONTV-A tips (10 nm nominal tip radius).

### 2.5. HR-EBSD
Residual elastic lattice strain and lattice rotation tensors were experimentally measured using the high angular resolution EBSD (HR-EBSD) technique (Wilkinson et al., 2006). A Kikuchi diffraction pattern was collected for each point at a resolution of 600 x 800 pixels using a conventional EBSD setup. EBSD experiments were performed in a Zeiss Merlin field emission gun SEM. Accelerating voltage of 20 keV and beam current of 15 nA were chosen. A Bruker e-flash high definition EBSD detector was used to collect EBSD patterns The EBSD maps were acquired using 169 nm step size. The maps were analysed using the XEBSD code (provided by A. J. Wilkinson (Wilkinson et al., 2006)) to probe the distortion of diffraction patterns with respect to a reference pattern chosen from a nominally strain free region within the same grain. The measured distortions can be linked to the elastic deformation gradient, $\boldsymbol{F}^e$, i.e. the elastic component of the two-point tensor, that maps the undeformed state to the deformed state (Suchandrima Das et al., 2018). From $\boldsymbol{F}^e$, the elastic strain ($\boldsymbol{\varepsilon}^e$) and rotation ($\boldsymbol{\omega}^e$) are determined. To ensure reliable measurements, correlation with 40 regions of interest was used. Since a lattice dilatation will not cause a distortion of the Kikuchi pattern, the lattice strain, $\boldsymbol{\varepsilon}^e$, measured by HR-EBSD only contains deviatoric components. Further details about the HR-EBSD technique can be found elsewhere (Britton and Hickey, 2018; Pantleon, 2008; Wilkinson, 1996; Wilkinson et al., 2006).

## 3. Experiments to guide CPFE formulation

Initial experiments were conducted on grains of three different orientations in the unimplanted and the 1 dpa samples, to develop a hypothesis for the underlying defect-dislocation interaction on which the crystal plasticity formulation might be based. These two extremes were chosen as they should most prominently show the differences brought about by self-ion implantation.

Figure 2(a) shows AFM micrographs of indents in <001>, <011> and <111> grains in the unimplanted and 1 dpa sample. While little difference is seen between indents in the unimplanted sample, a clear orientation dependence is observed in the 1 dpa sample. The <001> oriented grain shows substantial pile-up, while the other two orientations show little pile-up. This observation has two important implications:

First it may explain the seemingly contradictory indent surface morphologies reported in different self-ion implanted materials. Indents in self-ion implanted W-5wt%Ta showed suppression of pile-up (Armstrong et al., 2011), while a pile-up increase was seen around indents in self-ion implanted Fe-12wt% Cr (Hardie et al., 2015). Neither study mentions the crystallographic orientation of the grain under investigation. As such it is quite possible that the reported marked differences are simply the result of different grain orientations being probed (e.g. near <001> orientation in Fe-12 wt%Cr (Hardie et al., 2015) and a near <011>, <111> or an in-between orientation in W-5wt% Ta (Armstrong et al., 2011)).

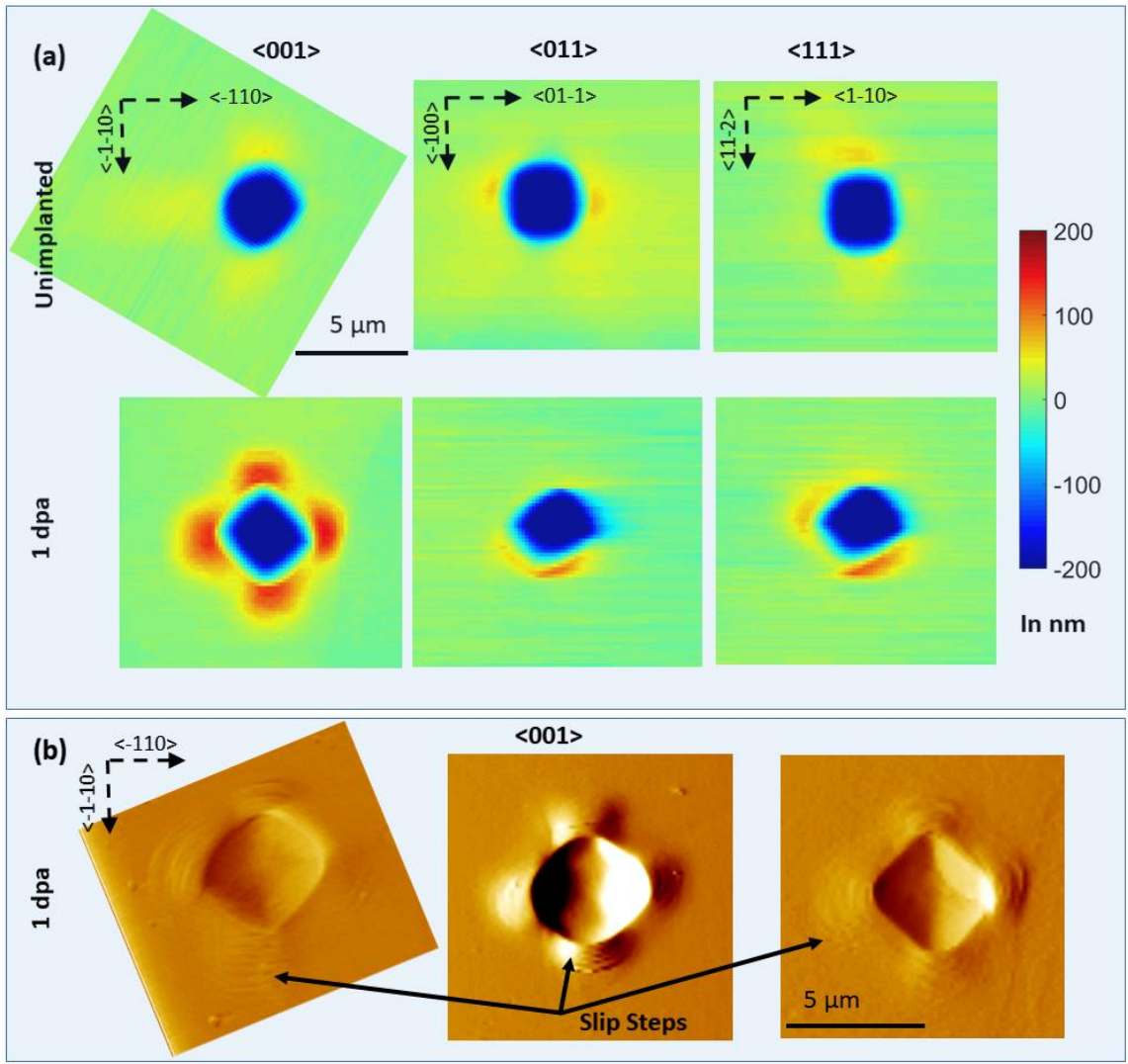

*Figure 2 – (a) AFM micrographs of 500 nm deep indents in grains of three different orientations (<001>, <011> and <111>) in the unimplanted and 1 dpa self-ion implanted tungsten samples. For each orientation, the micrographs in both samples have been rotated to maintain the same in-plane orientations. (b) Magnified view of AFM gradient images of indents in three <001> grains in the 1 dpa sample clearly showing the formation of slip steps.*

The second key implication is derived from the surprising likeness between the orientation-dependent pile-up patterns noticed here and those observed in 3000 appm helium-implanted tungsten (W-3000He)[1] (Das et al., 2019a). In particular, the localisation of pile-up and slip steps around the <001> indents in the 1 dpa sample (Figure 2b) closely match the observations from indents in <001>-oriented grains of W-3000He.

---

[1] In helium-implanted tungsten too, a large pile-up was seen for indents in <001> grain orientation and very little pile-up for indents in the <011> and <111> grain orientations.

In W-3000He, the mechanism of deformation in the presence of implantation defects, was found to be orientation-independent, and a physically-based CPFE formulation, implementing strain-softening, could reproduce the varying pile-up pattern for all orientations (Das et al., 2019a). The formulation was based on observations from several different experimental studies, as well as multi-scale simulations. Ab-initio calculations and lattice strain measurements suggest a damage microstructure dominated by helium-filled Frenkel-pairs (Becquart and Domain, 2009; Hofmann et al., 2015b). Based on the observations of increased hardening, slip traces near indents (observed by SEM) and reduced defect density channels beneath indents (observed by TEM), it was hypothesized that helium-filled Frenkel defects initially act as effective obstacles to glide dislocations. However, their obstacle strength is reduced by the passage of dislocations (probably due to glide-dislocation-assisted recombination of some of the Frenkel defects), leading to a localisation of deformation in slip channels (Das et al., 2019b).

Considering the remarkable similarity in observations from nano-indentation of self-ion- and helium- implanted tungsten (large pile-up and slip traces around <001> indents, reduced pile-up around <011> and <111> indents and increased hardening (Armstrong et al., 2013; Das et al., 2019a)), we propose that, although the defect microstructure induced by the two implantation conditions is distinctly different, orientation-independent strain-softening occurs during deformation in both cases. We hypothesize that in self-ion implanted materials too, the defect-dislocation interaction mechanism is independent of the orientation of the grain (relative to the surface or the indenter). Rather differences in surface morphology arise simply as a result of the relative orientation between the crystal (the specific grain in question), the sample surface and the spherical nano-indenter (Das et al., 2019a).

Our hypothesis for irradiation-induced strain softening in self-ion implanted tungsten is further strengthened by multiple previous experimental observations of strain localisation through defect-dislocation interactions, defect clearing and slip line or channel formation, in irradiated materials (Byun and Hashimoto, 2006; Farrell et al., 2004; Makin and Sharp, 1965; Zinkle and Singh, 2006). Besides our past work on CPFE modelling of strain-softening in helium-implanted tungsten, several other studies have reported modelling efforts to capture strain localisation in irradiated metals. For example, Rubla et al. used molecular dynamics (MD) and discrete dislocation dynamics (DD) to demonstrate dislocation pinning by irradiation-induced defects (such as vacancy stacking-fault tetrahedral in irradiated copper and self-interstitial atom Frank sessile loops in irradiated palladium), unpinning of the dislocations with increasing stress, absorption of the defects and formation of channels (Diaz De La Rubla et al., 2000). The combination of MD and DD has been subsequently used to explore dislocation channelling in irradiated fcc alloys, such as CuCrZr and CuNiBe (Gururaj et al., 2015; Singh et al., 2002). Crystal plasticity models, based on the theory of strain localisation, have also been reported for irradiated bcc metals such as iron (Barton et al., 2013; Patra and McDowell, 2012; Xiao et al., 2015).

Here we develop a physically-based CPFE material model for self-ion implanted tungsten of different doses, by modifying our prior CPFE formulation developed for W-3000He. The primary modification involves replacing the interaction of glide dislocations with helium-filled Frenkel pairs with a model that explicitly accounts for interaction of glide dislocations with irradiation-induced dislocation loops. Information about the density and size distribution of the dislocation loops at varying doses is obtained from TEM and directly linked to the model parameters. Simulations of the nano-indentation process are performed for a number of different self-ion damage levels, concentrating on grains with <001> orientation, where pile-up is expected to be highest. To examine the accuracy of the formulation, its predictions at varying dose levels and for varying crystallographic orientations are directly compared to AFM and HR-EBSD measurements.

## 4. CPFE Formulation

The nano-indentation experiments were simulated using a strain-gradient crystal plasticity model where dislocation slip was restricted to occur in slip directions compatible with the crystallography. The finite element software Abaqus 2016 (Dassault Systèmes, Providence, RI, USA), was used to simulate the indentation process. The crystal plasticity model was implemented in an Abaqus user material subroutine (UMAT). The model was constructed as having <001> crystallographic orientation, similar to the physical sample.

### 4.1. The CPFE Model

A 3D model comprising of a 20×20×20 $\mu m^3$ deformable block and a rigid 5 µm radius spherical indenter was constructed (Figure 1(b)). Considering symmetry for the <001> grain orientation, only a quarter of the experimental setup was modelled. Symmetry boundary conditions were applied on the XZ and YZ planes. The top surface was traction free and all other surfaces were fixed. The 20 µm high sample block was partitioned into two layers: a 2.5 µm thick implanted surface layer and a 17.5 µm thick substrate (Figure 1 (b)). In our prior study on helium-implanted tungsten (Das et al., 2019b) we found that the underlying substrate can have a significant effect on the results, even when the indentation depth is less than 20% of the implanted layer thickness. Thus, to ensure fidelity in our simulations, we account for the combined effect of the implanted layer and the substrate, instead of considering the implanted layer as an infinite half space.

The indenter was considered a discrete rigid wire frame while the sample block was modelled with isotropic elastic properties for tungsten (values in Appendix C). A frictionless hard contact was assumed between the indenter and the sample surface, as the mechanical response has been found to be insensitive to friction coefficient (Wang et al., 2004). As in the experiments, displacement-controlled loading was simulated where the indenter was subjected to a displacement of 0.5 µm into the sample block before unloading. The simulated load was scaled with an effective modulus $E_{eff}$ to account for the indenter tip compliance (Appendix B). The sample block was meshed using 20-node 3D quadratic hexahedral elements, with reduced integration (8 integration points) (C3D20R). A refined mesh (applied edge bias 0.1 to 2 µm) with 39500 elements was used (Figure 1 (b)).

### 4.2. UMAT formulation for pure tungsten

The user element (UEL) originally developed by Dunne et al. (Dunne et al., 2007) forms the basis of the CPFE implementation. A detailed description can be found elsewhere (Suchandrima Das et al., 2018; Dunne et al., 2007). Briefly, the slip law used is physically-based and considers the thermally activated glide of dislocations in a field of pinning dislocations. When the yield criterion is satisfied i.e. the resolved shear stress on slip system $\lambda$, $\tau^\lambda$, is greater than the critically resolved shear stress (CRSS), $\tau_c$, the crystallographic slip rate $\dot{\beta}_p^\lambda$ for slip system $\lambda$ is given by

$$\dot{\beta}_p^\lambda(\tau^\lambda) = \rho_m b^2 v \exp\left(-\frac{\Delta F}{kT}\right) \sinh\left(\frac{(|\tau^\lambda| - \tau_c)V}{kT}\right) sgn(\tau^\lambda) \qquad (1)$$

where, $\rho_m$, is the density of mobile dislocations, $v$ the attempt frequency, $b$, the Burgers' vector magnitude, $\Delta F$, the activation energy, $k$ the Boltzmann constant, $T$ the absolute temperature, and $V$ the activation volume, which depends on the spacing between the pinning dislocations, $l$. $l$ is estimated as $\frac{1}{\sqrt{\Psi(\rho_{SSD})}}$, where, coefficient $\Psi$ represents the probability of pinning by $\rho_{SSD}$ i.e. the density of statistically stored dislocations (SSD). The values of the material parameters used here are provided in Appendix C. Assumptions of isotropic elasticity and small elastic deformations were made in the formulation. Also, for simplicity, all parameters on the RHS in Eq. (1) were kept unchanged, other than $\tau_c$ (considered the same for all slip-systems) and the independent variable $\tau^\lambda$.

The critically resolved shear stress (CRSS) at any point in the material is

$$\tau_c = \tau_c^0 + C'G\,b\,\sqrt{\rho_{GND}}. \qquad (2)$$

The first term on the RHS of Eq. (2), $\tau_c^0$, is the CRSS of the pure unimplanted tungsten. As the material deforms plastically, new dislocations are created to accommodate the lattice curvature, i.e. geometrically necessary dislocations (GNDs) (Suchandrima Das et al., 2018; Nye, 1953). The evolution of these GNDs increases the number of obstacles encountered by gliding dislocations. The second term accounts for this strain hardening using a Taylor hardening law (Davoudi and Vlassak, 2018; Taylor, 1934) where $C'$ is a hardening factor, $G$ the shear modulus of tungsten and $\rho_{GND}$ the sum of the GNDs produced across all slip-systems. Only two parameters were fitted to the experimental results (nano-indentation and AFM surface profile) of the unimplanted sample; $\tau_c^0$, and $C'$.

For simulating indentation of pristine tungsten, both layers of the model were assigned the properties of the unimplanted material. When simulating indentation of the self-ion implanted tungsten sample, material parameters for the undamaged substrate layer were kept unaltered, while additional features were added to the top layer representing the self-ion implanted tungsten (Section 4.3).

## 4.3. UMAT formulation of self-ion implanted tungsten

Based on the observations from the 1 dpa implanted sample (Section 3), the UMAT formulation for self-ion implanted tungsten is built on the same strain-softening formulation as used for helium-implanted tungsten (Das et al., 2019b). The hypothesis is that initially the self-ion induced loops pose strong obstacles to gliding dislocations, causing hardening. However, their strength is reduced by glide dislocations cutting through them (details in Section 4.3.1). This leads to the localisation of deformation in channels and hence the formation of slip steps.

To implement this hypothesis, Eq. (2) is modified for the implanted layer by including an extra term accounting for the additional shear resistance $\tau_H$ (with initial value $\tau_H^0$ at *t = 0*) of the implantation-defects

$$\tau_c = \tau_c^0 + C'G\,b\,\sqrt{\rho_{GND}} + \tau_H. \tag{3}$$

To implement strain-softening, i.e. the progressive weakening of defects by gliding dislocations, $\tau_H$ is reduced at the end of each time increment $\Delta t$, as a function of the accumulated crystallographic slip, $\beta_p$:

$$\beta_p^{t+\Delta t} = \beta_p^t + \sum_{\lambda=1}^{n} \dot{\beta}_p^\lambda \Delta t \tag{4}$$

$$\tau_H^{t+\Delta t} = \tau_H^0 e^{-(\beta_p^{t+\Delta t}/\gamma)} \tag{5}$$

where, $\beta_p^t$ and $\beta_p^{t+\Delta t}$ are the accumulated slip summed over all slip systems at the start and end of increment $\Delta t$. The rate of defect removal is likely to be proportional to its current value i.e. the current defect concentration[2] (Das et al., 2019b) resulting in the exponential softening rate in Eq. (5).

---

[2] $\left.\frac{\partial \tau_H}{\partial \beta_p}\right|_{t+\Delta t} = -\tau_H/\gamma$

Thus, two new parameters are introduced in the implanted layer: $\tau_H^0$ and the softening rate $\gamma$ (Eq. (5)). Of these, only $\gamma$ is fitted to the experimental data of the 0.01 dpa implanted sample and is kept constant in simulations for all other damage levels. $\tau_H^0$ is physically-based and derived from TEM data of defect density for the different damage levels as described in Section 4.3.1

### 4.3.1. Determining $\tau_H^0$

$\tau_H^0$, the initial value of the implantation-induced shear resistance force, is computed specifically for each damage level based on the implantation-induced loop number density determined by TEM as a function of damage (Yi et al., 2016). While TEM uniquely provides a direct image of the irradiation-induced defects, there remain some challenges associated with it. TEM data alone may not fully capture the complete population of irradiation-induced damage. In particular it has been shown to lack sensitivity to small defects (<1.5 nm) (Zhou et al., 2006). These may be probed using complementary techniques such as positron annihilation spectroscopy (PAS) (for capturing vacancy-related defects) (Barthe et al., 2007; Debelle et al., 2008; Wiktor et al., 2014), X-ray micro-beam Laue diffraction (for measuring defect-induced strains) (S. Das et al., 2018b; Hofmann et al., 2015b) or elastic recoil detection analysis (ERDA) (for resolving the depth-dependent variation of defect population) (Grigull et al., 1997; Siketić et al., 2018). While establishing the complementarity of these experimental techniques for extracting a complete picture of irradiation-induced damage is very important, it is beyond the scope of this study. Here our aim is to demonstrate the utility of experimental techniques, such as TEM, in providing a physical basis for the parameters used in numerical formulations. Thus, we make the following assumptions in using the TEM data to establish $\tau_H^0$.

Loops smaller than 1.5 nm, i.e. below the sensitivity limit of TEM, are assumed to have little effect on dislocation glide compared to larger loops. Loop loss to surfaces due to image forces, which may occur in TEM foil samples, is not accounted for. Further, we assume that the size distribution and density of loops, as captured by TEM, is representative of and applicable to the whole thickness of the implanted layer. As noted in Section 2.4, there is a small variation in the induced damage across the thickness of the implanted layer (standard deviation of 0.23 dpa for the 1 dpa implanted sample or ~20%) (Figure 1(a)). However, our previous thermal transport measurements on the same material system suggest that the differences in damage microstructure that arise due to this variation in dose are small (Reza et al., 2020a). Thus, a uniform damage distribution throughout the implanted layer thickness, is considered for the CPFE simulations.

Yi et al. (Yi et al., 2016) used TEM to study tungsten samples implanted with 150 keV $W^+$ ions at 300 K for a range of damage levels from 0.01 up to 1 dpa. Though the ion implantation energy in the referred study is different from that used here, the data is relevant because of the comparable dpa levels.

We start by considering the TEM observations for a damage level of 0.01 dpa. At this early stage of damage there is negligible cascade overlap (Yi et al., 2016)). For this case, TEM observations detailing the frequency of occurrence ($f$) of a loop containing $N$ point defects per implanted ion (Figure 3 in (Yi et al., 2016)) can be determined (extracted data points in Table 2, columns 1 and 2 and plotted in (Figure 3)). From the data in Figure 3, $L_d^N$ i.e. the number density of loops (in loops/m²) of diameter $d$ and containing $N$ point defects within the implanted layer of the TEM foil can be calculated. The loop diameter $d$ can be computed from $N$ as

$$d = \sqrt{N} \frac{2a}{3^{1/4}\sqrt{\pi}} \qquad (6)$$

where, $a$ is the lattice parameter (0.31652 nm for tungsten (Derlet et al., 2007; Hofmann et al., 2015b)). Eq. (6) is derived based on the assumption that all visible defects are ½<111> circular prismatic dislocation loops with area $A = \pi d^2/4$ containing $N = Ab/V$ atoms, with atomic volume, $V = a^3/2$, and Burgers' vector length, $b = \sqrt{3}a/2$, for bcc tungsten (Yi et al., 2016).

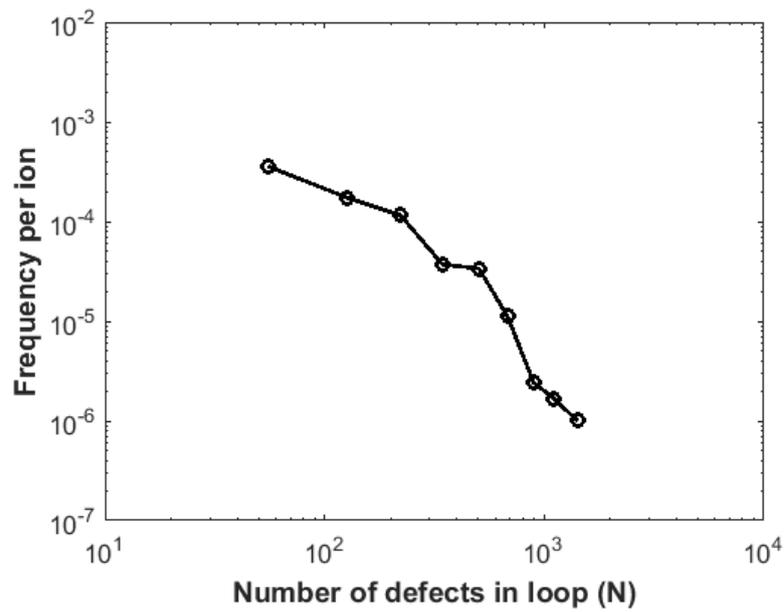

*Figure 3 – Plot of data points extracted from (Yi et al., 2016) showing the frequency of occurrence (f) of a loop containing N point defects for every implanted ion for tungsten implanted with self-ions at 300 K to 0.01 dpa.*

In Yi et al. (Yi et al., 2016) the bins in the plots are defined as $[N(d-0.5), N(d+0.5)]$ ($d$ being the loop diameter in nm) and $f$ is normalized to the bin width and the ion-fluence $\varphi$ (which for 0.01 dpa is $1 \times 10^{16}$ ions/m²). Thus, for a given value of $N$ from the plot, we can use Eq. (6) to compute the corresponding loop diameter, $d$, and thereby the corresponding bin width, $B_w$, as

$$B_w = N(d+0.5) - N(d-0.5) \qquad (7)$$

Knowing $\varphi$, we can compute the loop number density $L_d^N$

$$L_d^N = f\varphi B_w. \tag{8}$$

$L_d^N$ here is the loop number density (in loops/m²) for the whole thickness of the irradiated layer in the TEM foil, which in this case is $D = 25$ nm (the total foil thickness is reported as 66 nm (Yi et al., 2016) and the irradiated layer thickness is ~ 25 nm). Here we assume that the loops are uniformly stacked with zero vertical spacing throughout the 25 nm thickness of the irradiated foil (Figure 4 (a)). Thus, as a dislocation glides through a planar section of the irradiated layer (Figure 4 (b)), it will encounter only a certain proportion of $L_d^N$, which we refer to here as $\rho_d^N$ and compute as

$$\rho_d^N = \frac{L_d^N}{\left(\dfrac{D}{d}\right)} \tag{9}$$

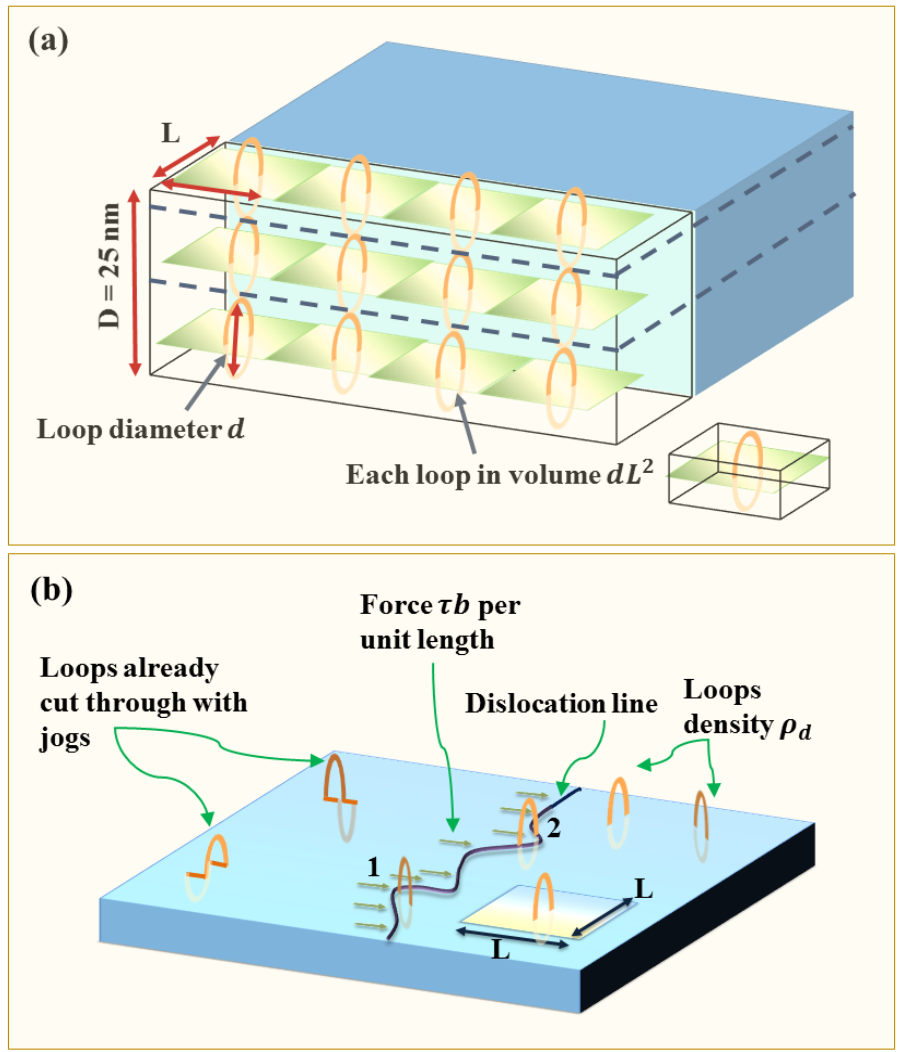

Figure 4 – (a) Schematic of ion-implanted TEM foil cross-section (thickness D = 25 nm, the thickness of the implanted layer in Yi et al. (Yi et al., 2016)). Depicted are the assumptions made in the model of uniformly stacked loops with zero vertical spacing. (b) Illustration of dislocation glide through a planar section of the implanted layer, enclosed by the dashed line in (a). The glide plane is interspersed with implantation-induced loops (only few loops in the plane are illustrated here) that act as obstacles to the motion of the glide dislocation. Positions 1 and 2 show the glide dislocation interacting with a loop edge-on and face-on respectively.

Table 2– Data extracted from TEM investigation of tungsten implanted tungsten of damage level 0.01 dpa (Yi et al., 2016) showing the density of loops of various sizes and the possibility of their encounter with gliding dislocations.

| Number of point defects in loop ($N$) | Frequency of creation of loop ($f$) | $L_d^N$ (loops/m²) | $\rho_d^N$ (loops/m²) |
|---|---|---|---|
| 55.1 | 0.000359 | $1.96 \times 10^{14}$ | $1.58 \times 10^{13}$ |
| 126 | 0.000174 | $1.44 \times 10^{14}$ | $1.75 \times 10^{13}$ |
| 222 | 0.000115 | $1.26 \times 10^{14}$ | $2.04 \times 10^{13}$ |
| 344 | $3.72 \times 10^{-5}$ | $5.09 \times 10^{13}$ | $1.02 \times 10^{13}$ |
| 504 | $3.38 \times 10^{-5}$ | $5.59 \times 10^{13}$ | $1.36 \times 10^{13}$ |
| 679 | $1.13 \times 10^{-5}$ | $2.17 \times 10^{13}$ | $6.14 \times 10^{12}$ |
| 895 | $2.42 \times 10^{-6}$ | $5.34 \times 10^{12}$ | $1.73 \times 10^{12}$ |
| 1110 | $1.66 \times 10^{-6}$ | $4.08 \times 10^{12}$ | $1.47 \times 10^{12}$ |
| 1420 | $1.00 \times 10^{-6}$ | $2.78 \times 10^{12}$ | $1.14 \times 10^{12}$ |
| | | $L_{tot} = \sum L_d^N = \mathbf{6.07 \times 10^{14}}$ | $\rho_{cut} = \sum \rho_d^N = \mathbf{8.81 \times 10^{13}}$ |

For each extracted data point from Figure 3, the calculated $L_d^N$ and $\rho_d^N$ are shown in Table 2. $L_d^N$ summed over the range of loop sizes gives the total loop number density ($L_{tot}$) in the 25 nm thick implanted layer of the TEM foil (Table 2). $\sum \rho_d^N$ summed over the range of loop sizes gives $\rho_{cut}$, the number density of loops that will interact with a dislocation gliding through a plane in the implanted layer of the foil (Table 2), and that will contribute to $\tau_H^0$ (i.e. the increased shear resistance of the implanted material).

We note here the ratio between $\rho_{cut}$ and $L_{tot}$ (Table 2) is $P = \rho_{cut}/L_{tot} = 0.0145$ (This ratio is specific to the 150 keV ion energy irradiated 0.01 dpa ion-implanted sample reported by Yi et al. (Yi et al., 2016)). We use this as a conversion factor to compute $\rho_{cut}$ the loop number density intersecting a slip plane from the reported total loop number density ($L_{tot}$) for the other damage levels. While defect morphology may change with increasing dose, detailed knowledge of these changes is difficult to ascertain. Thus, in order to minimise assumptions and fitting parameters in the model, we assume that $P$ remains constant across all damage levels. This assumption implies that there is a linear scaling of the number density of loops of certain diameter (i.e. $L_d^N$ for loops of diameter $d$ and containing $N$ point defects) with the total number density (i.e. $L_{tot}$) of loops, irrespective of the damage level.

The total loop number density ($L_{tot}$) has been reported by Yi et al. for a range of damage levels (Figure 7a in (Yi et al., 2016) ). The reported values of $L_{tot}$ for the damage levels 0.1, 0.32 and 1 dpa were extracted (Table 3). $L_{tot}$ for 0.01 dpa is already known from Table 2. The corresponding $\rho_{cut}$ is calculated as $\rho_{cut} = PL_{tot}$.

Table 3 – The implantation-induced additional shear resistance force to dislocation glide $\tau_H^0$ computed for the different damage levels based on the defect density of loops as estimated from TEM of the damage microstructure in 150 eV W⁺ implanted tungsten by Yi et al. (Yi et al., 2016).

| Damage Level (dpa) | $L_{tot}$ (loops/m²) | $\rho_{cut}$ (loops/m²) | $\tau_H^0$ (MPa) |
|---|---|---|---|
| 0.01 | $6.07 \times 10^{14}$ | $8.81 \times 10^{13}$ | 260 |
| 0.1 | $3.16228 \times 10^{15}$ | $4.59352 \times 10^{14}$ | 588 |
| 0.32 | $4.21697 \times 10^{15}$ | $6.12556 \times 10^{14}$ | 679 |
| 1 | $5.37032 \times 10^{15}$ | $7.80092 \times 10^{14}$ | 766 |

To derive $\tau_H^0$ as a function of $\rho_{cut}$, a segment of a gliding dislocation of length $L$ is considered to be moving through the planar section which is intersected by $\rho_{cut}$ loops/m² (Figure 4(b)). Each loop is considered to be occupying a planar area of $L^2$ i.e. $\rho_{cut} = 1/L^2$.

As the gliding dislocation cuts through the "forest" of implantation-induced loops, it displaces material above the slip plane, creating jogs in the cut dislocation loops. A model is considered here where the gliding dislocation may encounter a loop edge-on (Figure 4(b) position 1) or face-on (Figure 4(b) position 2), or in a position in between these two.

We note here that the mechanism of interaction between dislocations and loops may change as a function of the loop diameter. In future, equations accounting for such detailed interactions can be developed through dislocation dynamics studies of a progressively evolving loop-dominated microstructure. For simplicity, here we assume the same mechanism of interaction between dislocations and loops of all sizes and the effect of loop size is accounted for in the computation of $\rho_{cut}$ (Table 2).

For edge-on loop interactions, one jog is created (equivalent to cutting through one dislocation in a dislocation forest). The length of the jog is equal to the Burgers' vector magnitude $b$. Thus, the total energy required for the dislocation to cut through is equal to the energy needed to create the jog i.e. $Gb^2b/2$, where $G$ is the shear modulus. The force driving the dislocation segment forward is $\tau_H^0 bL$. Thus the work done in moving the dislocation forward by a distance $b$ is $(\tau_H^0 bL)b$. Equating energy with the work done gives

$$\tau_H^0 b^2 L = Gb^3/2 \qquad (10)$$

Therefore,

$$\tau_H^0 = \frac{Gb}{2L} \qquad (11)$$

Eq. (12) can be re-written by expressing $L$ in terms of $\rho_{cut}$

$$\tau_H^0 = \frac{Gb}{2}\sqrt{\rho_{cut}} \qquad (12)$$

Eq. (13) computes $\tau_H^0$ for edge-on loop encounter where one jog is created at a time. In the case of a face-on loop interactions, the dislocation will have to create two jogs simultaneously to cut through the loop, i.e. in this case $\tau_H^0 = 2\frac{Gb}{2}\sqrt{\rho_{cut}}$. For small loops with $r \ll L$ this may be a more accurate approximation. In the general case of loop encounter with a dislocation, we can then re-write Eq. (13) as

$$\tau_H^0 = m\frac{Gb}{2}\sqrt{\rho_{cut}} \qquad (13)$$

where $1 \leq m \leq 2$ is a pre-factor that accounts for the face-on or edge-on nature of dislocation-loop interaction. In other words, $m$ represents the average number of jogs created simultaneously by the dislocation as it cuts through the pinning loops. $m = 1.23$ was obtained from fitting to the experimental result of the 0.01 dpa sample and kept constant for simulations of the other damage levels. Using this and the respective $\rho_{cut}$ values for each damage level as computed in Table 3, $\tau_H^0$ for each damage level is computed (Table 3).

## 5. Results & Discussion

For purposes of direct comparison, the simulation and experimental results are plotted in the same coordinate frame and using the same colour and length scales.

### 5.1. Mechanical response from nano-indentation

Figure 5 (a) shows the measured load-displacement curves from nano-indentation of <001> oriented grains in the self-ion implanted tungsten samples exposed to different damage levels. One measurement was made in the unimplanted sample. Each of the other curves is the average of three different indents in each sample. The initial Hertzian elastic response in the curves in Figure 5 (a) is almost identical in the unimplanted and implanted samples. This behaviour is expected since ion-implantation-induced changes in elastic modulus are small (Duncan et al., 2016; Hofmann et al., 2015b). The unimplanted sample, shows a large pop-in event at ~60 nm penetration depth. This is indicative of the onset of plastic deformation and associated nucleation of dislocations (Ma et al., 2012). However, no such obvious pop-ins are observed in any of the implanted samples. This implies that while the unimplanted material is relatively defect free, the implanted samples, populated by implantation-induced defects, allow an easier nucleation of initial glide dislocations.

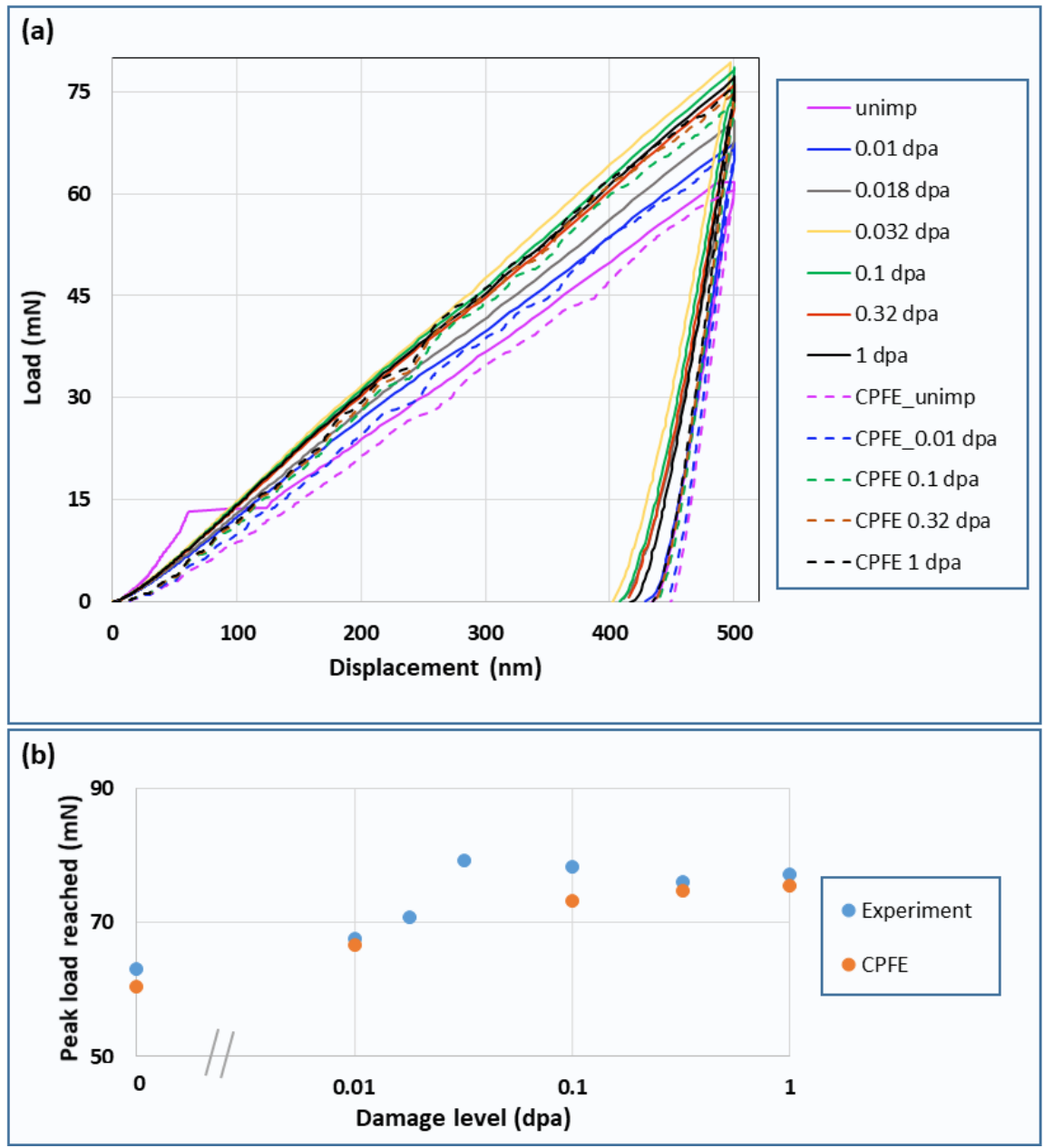

*Figure 5 – (a) Load-displacement curves as measured by nano-indentation and predicted by CPFE for <001>-oriented grains in self-ion implanted samples with a range of damage levels (0 – 1 dpa). The solid curves are the average of three different indents in each sample. (b) Plot of peak load reached by the ion-implanted samples with different damage levels as measured by nano-indentation and predicted by CPFE. We note that CPFE simulates indentation for only 5 damage levels, excluding 0.018 and 0.032 dpa.*

With increasing indentation depth, the implanted samples reach a noticeably higher load than the unimplanted material. This is consistent with the well-known hardening effect induced by irradiation defects, which has been previously reported in tungsten as well as other self-ion implanted materials, e.g. iron and iron alloys (Armstrong et al., 2011; Gibson et al., 2015; Hardie et al., 2015). The early saturation of this hardness increase, on the other hand, is surprising. Tracing the increase in hardness with irradiation, the 0.01 dpa sample shows a ~9% increase, 0.018 dpa sample ~13% increase and the 0.032, 0.1, 0.32 and 1 dpa samples all show ~27% increase. The saturation of hardening as early as 0.032 dpa (Figure 5 (b)) is lower than previous observations of hardness saturating above 0.4 dpa in ion-implanted tungsten (implanted with self-ions at 300 °C) (Armstrong et al., 2013).

The saturation of hardness is also reflected in the morphology of the surface pile-up around indents. AFM was used to measure the indent morphologies in the unimplanted sample, the 0.01 dpa sample and three samples above the saturation threshold level, 0.1, 0.32 and 1 dpa. To ensure reproducibility of results, three indents in <001> grains were measured for each sample. Figure D.1 in Appendix D shows that the results are consistent. Figure 6 shows the AFM measurements for one of the indents from each sample. The indent in the unimplanted sample shows a shallow pile-up that extends quite far from the indent (up to about 6 μm), visible in the form of pile-up streaks along the {110} directions. The pile-up around the indents becomes more confined and increases in height as implantation damage increases. Beyond 0.1 dpa, there is little change in the pile-up profile. Increase in pile-up with damage level and saturation beyond 0.1 dpa is clearly demonstrated by line plots made through the AFM measurements (Figure 6) along the <110> direction as shown in Figure D.2(a) in Appendix D.

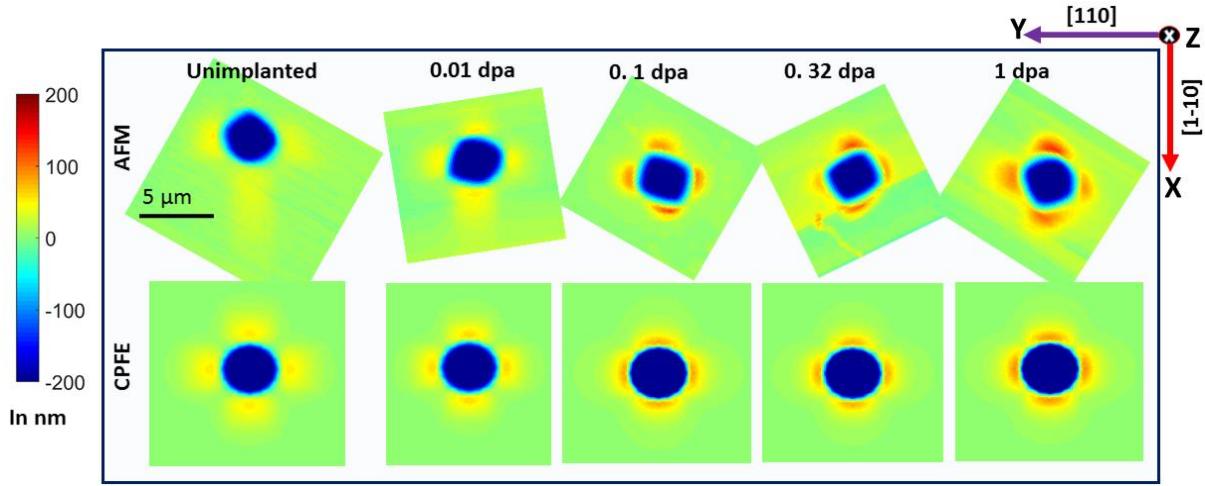

*Figure 6 – Surface morphologies (height profile) of nano-indents in pure tungsten and in self-ion implanted tungsten with varying damage levels as predicted by CPFE and measured by AFM after indentation. The AFM micrographs have been rotated to have the same in-plane orientations as the CPFE model. The surface normal pointing out of the page is along [00-1]. The colour scale shows surface height in nm.*

This saturation effect implies a saturation of the underlying defect population. A similar saturation in defect densities was found through TEM studies on self-ion implanted tungsten, tungsten-5%-rhenium and tungsten-5%-tantalum (all implanted at 150 keV) beyond 0.4 dpa (Yi et al., 2016). It has been seen that with increase in dose > 0.4 dpa, loop dynamics change resulting in the formation of strings and larger finger loops. In the process, freshly nucleated loops are absorbed, stalling further defect accumulation (Yi et al., 2016; Yu et al., 2020). Similar observations of saturation of defect densities and evolution of defect morphologies beyond 0.1 dpa has also been noticed in 150 keV self-ion implanted iron and iron-chromium alloys (Yao et al., 2008).

Interestingly our observations suggest that, despite changing defect morphology in the saturation region, the overall effect of irradiation-induced defects on glide dislocations does not change significantly. Rather the material hardness and surface pile-up, and consequently dislocation mobility, remain largely the same across the higher damage levels.

To further explore this observation, CPFE was used to simulate nano-indentation for the unimplanted tungsten, 0.01 dpa sample and three damage levels at and above the saturation level, 0.1, 0.32 and 1 dpa. The surface profiles predicted by CPFE are compared with the AFM measurements (Figure 6). Very good agreement is observed between the CPFE and AFM measurements across all the samples. The confinement of deformation along the slip directions consistent with the crystallography of bcc tungsten is clearly captured, leading to the expected four-fold pile-up pattern (Wang et al., 2004). CPFE also reproduces the significant confinement of pile-up and increase in pile-up height for damage levels (> 0.1 dpa), in good agreement with the experimental results. Good quantitative agreement of the surface pile-up predicted by CPFE with the corresponding experimental observations, for each investigated dose, is highlighted by superimposition of line plots made along the <110> directions as shown in Figure D.2 (b)-(f).

The simulated load-displacement curves are superimposed as dotted lines on the respective nano-indentation measurements in Figure 5. The CPFE predictions capture the saturation in load-curves, with the 0.32 and 1 dpa samples showing ~ 3% difference and the 0.1 dpa sample, showing a 7% difference to the experimental observations. Quantitative agreement between CPFE predictions and experimental observations inspires confidence in the hypothesis of strain softening and the exponential rate of softening adopted in the model (Eq. (5)). The primary limitation of this model is that the underlying dynamics of dislocation nucleation and evolution are not accounted for. While this allows the model to have a minimal number of parameters, it renders it incapable of simulating features such as the pop-in in the unimplanted sample. However, despite this limitation, the model captures the implantation-induced changes in quite well.

Next, we use the CPFE formulation to investigate the orientation-dependent effects seen in Figure 2(a) in Section 3. In experiments, the implantation-induced increase in pile-up around the indent is exclusive to indent in <001> oriented grains. In Section 3, we hypothesized that, akin to helium-ion-implanted tungsten (Das et al., 2019a), in self-ion implanted materials too, the defect-dislocation interaction mechanism is orientation-independent. Rather differences in surface morphology arise simply as a result of the relative orientation between the crystal, the sample surface and the spherical nano-indenter (Das et al., 2019a). To explore the validity of this hypothesis, we simulated the indentation experiment for <111> and <011> orientations for the unimplanted and the 1 dpa implanted sample. For each considered case, the CPFE formulations and associated parameters were kept unchanged and only the input grain orientation was changed. Based on symmetry, for <011> grains, the model simulated one quarter of the experimental setup, using the mesh shown in Figure 1(b). For the <111> grain, which exhibits a one-third symmetry, a model (40×40×10 $\mu m^3$) simulating the whole experimental setup was used, as shown in Figure E.1 in Appendix E. The top surface of this model was kept traction free and all other surfaces were fixed. The sample block was meshed using 20-node 3D quadratic hexahedral elements, with reduced integration (8 integration points) (C3D20R). A refined mesh (applied edge bias 0.1 to 2 $\mu$m) with 77500 elements was used (Figure 1 (b)).

The AFM measurements shown in Figure 1(a) are re-plotted in Figure 7 (a)-(c) and (g)-(i) to allow a direct comparison with the corresponding predictions from CPFE (Figure 7 (d)-(f) and (j)-(l) respectively). The AFM micrographs in Figure 7 are rotated to have the same in-plane orientation as the profiles predicted by CPFE. As expected, four, two and three-fold symmetry can be seen in the CPFE predicted surface profiles of the <001>, <011> and <111>grains respectively. Figure 7 (j)-(l), clearly illustrates the orientation-dependence in pile-up pattern in the 1 dpa implanted sample as captured by the orientation-independent, underlying defect-dislocation interaction mechanism. In agreement with experiments, CPFE predicts, much lower pile-up in the <011> and the <111> grains than in the <001> grain. This agreement between the CPFE and experiments confirms our hypothesis that the deformation mechanism in self-ion-implanted samples, as in helium-ion-implanted samples, is orientation-independent. Further validation of the hypothesis as observed through comparison of the GND distribution for the three different orientation will be discussed in Section 5.2.

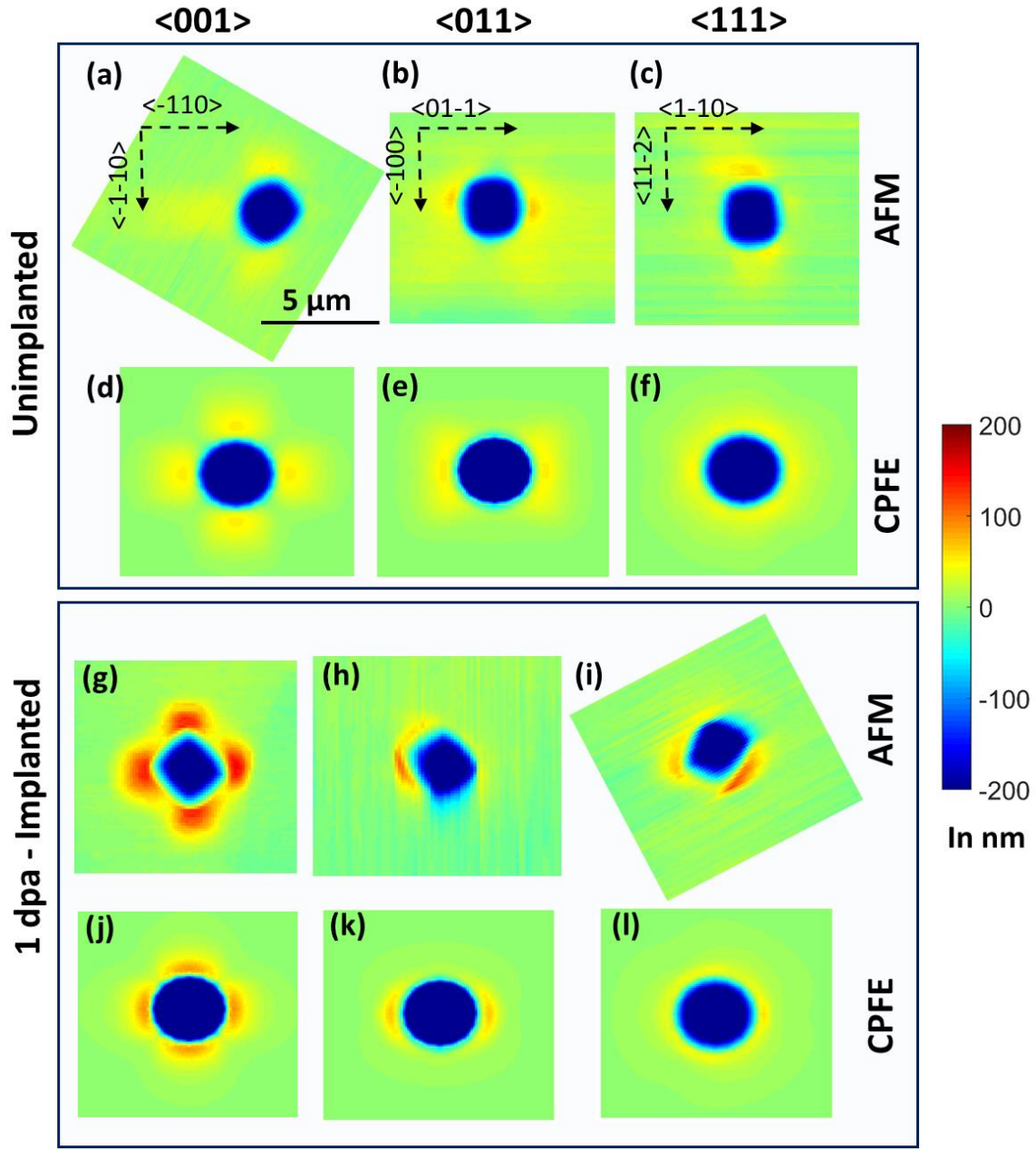

*Figure 7 - Surface profiles of residual out-of-plane displacement after indentation. AFM for the unimplanted (a) – (c) and the 1 dpa self-ion-implanted sample (g) – (i) for <001>, <011> and <111> out-of-plane crystal orientations respectively. CPFE simulations for the unimplanted (d) – (f) and the 1 dpa self-ion-implanted sample (j) – (l) for <001>, <011> and <111> out-of-plane crystal orientations respectively. The AFM micrographs have been rotated to match the in-plane orientations of the CPFE plots. The colour scale and 5 µm scale bar are the same for all plots. The in-plane orientations indicated for each orientation in (a)-(c) are applicable to all the remaining plots of each orientation. Although the indent depth below the surface, after unloading, is ~400 nm, a colour-scale of -200 to 200 nm is used as we concentrate on investigating pile-up morphology on the sample surface.*

## 5.2. Lattice distortions and GND density around indents

In addition to changes in indentation load-displacement curves and pile-up morphology, it is interesting to consider how irradiation-induced defects modify the lattice distortions and GND density distribution near indents. Here the two extremes, the unimplanted and the 1 dpa sample, are considered in detail. Lattice distortions around indents in both samples were probed using HR-EBSD. The experimental measurements are compared to the corresponding CPFE simulations, examining both residual lattice rotations and elastic lattice strains.

The indentation-induced change in orientation $\boldsymbol{R}$, for each point with final lattice orientation at time $t$, $\boldsymbol{R}_t$ was calculated as (Suchandrima Das et al., 2018):

$$\boldsymbol{R}_t = \boldsymbol{R}\boldsymbol{R}_0 \quad (14)$$

$$\boldsymbol{R} = \boldsymbol{R}_t \boldsymbol{R}_0^{-1} \quad (15)$$

where, $\boldsymbol{R}_0$ is the orientation of the un-deformed material. For CPFE, $\boldsymbol{R}_0$ was the original input orientation. For HR-EBSD $\boldsymbol{R}_0$ was taken as the average orientation of points far from the indent (25 μm away). Following the convention adopted in (Slabaugh, 1999), $\boldsymbol{R}$ captures the combined effect of sequential rotation about the X, Y and Z axis by the lattice rotation angles $\theta_x$, $\theta_y$ and $\theta_z$.

The deviatoric component of the residual elastic strain measured by HR-EBSD (Wilkinson, 1996) is compared with CPFE predictions. The deviatoric part of the CPFE predicted elastic strain ($\boldsymbol{\varepsilon}^e_{dev}$) was extracted as

$$\boldsymbol{\varepsilon}^e_{dev} = \boldsymbol{\varepsilon}^e - \boldsymbol{\varepsilon}^e_{vol} = \boldsymbol{\varepsilon}^e - 1/3 \, Tr\,(\boldsymbol{\varepsilon}^e)\boldsymbol{I} \quad (16)$$

where $\boldsymbol{\varepsilon}^e_{vol}$ is the volumetric strain.

Figure 8 (a) and (b) show the CPFE predictions of lattice rotations and all six components of the residual elastic deviatoric lattice strain for the unimplanted and the 1 dpa sample respectively. The corresponding measurements by HR-EBSD are shown in Figure 8 (c) and (d) respectively. We note that for the experimental measurements, in some instances, the indent may be located near a grain boundary (for e.g. Figure 8 (d)). To ascertain the impact of the grain boundary on the HR-EBSD measurements, we consider the lattice distortions measured around indents close to a grain boundary (e.g. measurements in the 1 dpa sample in Figure 8 (d)) with those made around an indent located well away from any grain boundary (e.g. measurements in the 0.32 dpa sample in Figure F.1 in Appendix F). The 0.32 dpa indent is comparable to the 1 dpa case as they are both in the saturation regime i.e. beyond 0.1 dpa, where irradiation-induced changes on the deformation behaviour are small, as shown in Figure 5 and Figure 6. The measurements made in the two cases are found to be almost identical. This confirms that the proximity of the indent to the grain boundary has no notable impact on the measurements of the lattice distortions and the subsequent GND density calculations.

The plots in Figure 8 show the XY plane i.e. the sample surface. HR-EBSD shows that in comparison to unimplanted tungsten, the magnitude of lattice strain and rotation is higher in the 1 dpa sample (particularly evident in $\varepsilon_{xx}$ and in $\varepsilon_{yy}$), but the fields are also spatially more confined. This confinement is particularly prominent in $\varepsilon_{xy}$, $\varepsilon_{zz}$ strain components and the lattice rotation fields. CPFE captures this localisation of the elastic fields well.

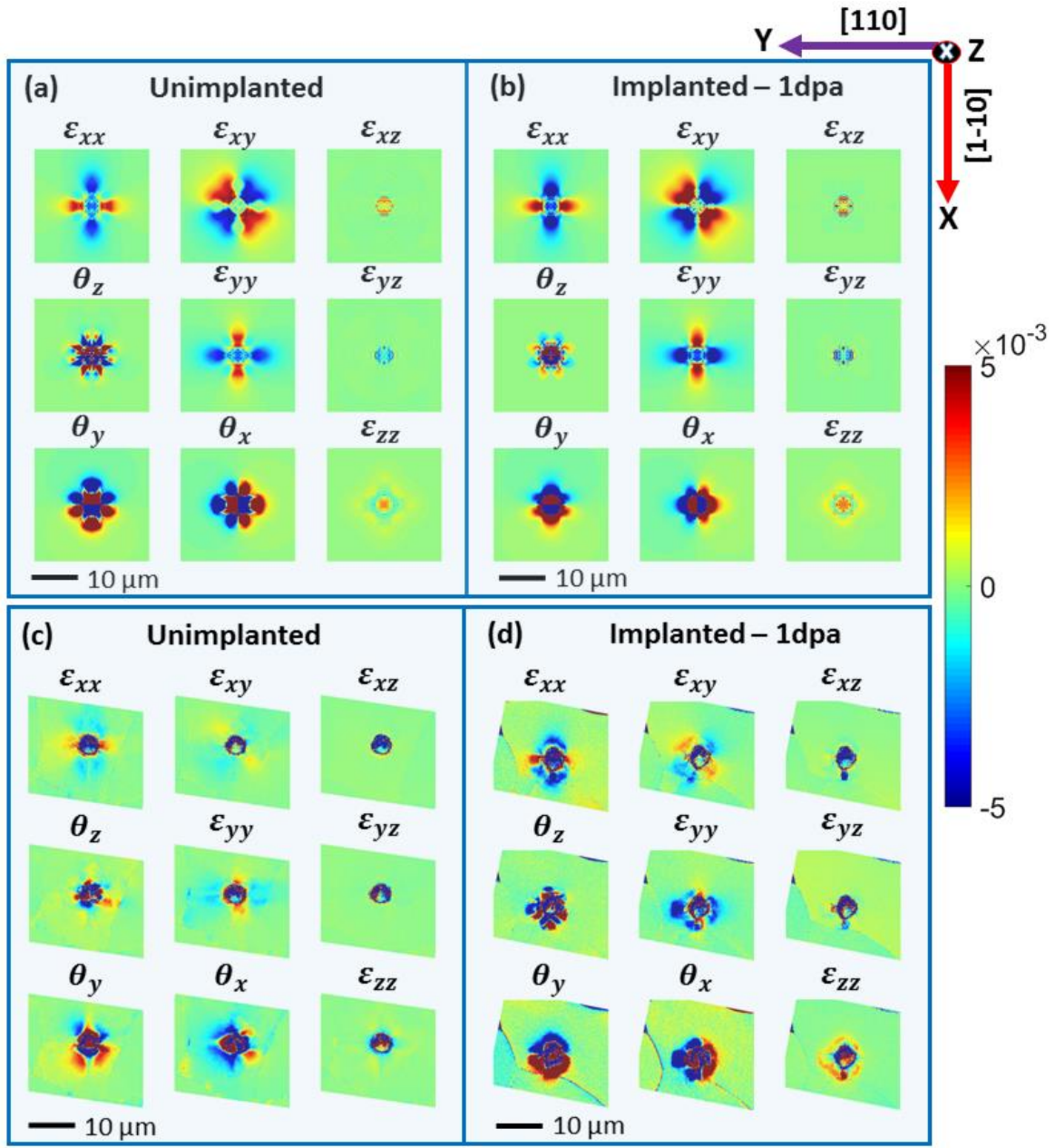

*Figure 8 —CPFE predictions of lattice rotations and all six components of the residual elastic deviatoric lattice strain plotted on the XY plane (sample surface) for the (a) unimplanted sample and (b) 1 dpa sample. HR-EBSD measurement of lattice rotations and all six components of the residual elastic deviatoric lattice strain plotted on the XY plane (indent surface) for the (c) unimplanted sample and (d) 1 dpa sample.*

CPFE predictions of implantation-induced change in magnitude and localisation of lattice distortions is particularly noticeable for the lattice rotations and strains $\varepsilon_{xy}$ and $\varepsilon_{zz}$. The pattern of negative and positive lobes for these components also agrees very well with HR-EBSD measurements. Shear strains are generally more difficult to measure than the direct components of strain (Villert et al., 2009). In this case however, HR-EBSD measurement and CPFE prediction of the in-plane shear strain $\varepsilon_{xy}$, is surprisingly consistent, including the similarity in irradiation-induced changes in magnitude and confinement. Little change is predicted by CPFE for the out-of-plane shear strains in both samples (which, as expected, are also much smaller than the other strain components), consistent with observations from HR-EBSD.

As lattice rotations are expected to play a dominant role in the overall deformation gradient (Nye, 1953), their distribution below the indents is also explored. Since HR-EBSD measurements are restricted to the indent surface, this is done exclusively through CPFE simulations. Figure 9 shows the lattice rotations beneath the indents, predicted by CPFE, for the unimplanted and the 1 dpa sample. They are plotted on virtual YZ slices at three points: At the indent centre (slice 2 in Figure 9) and 6 μm on either side of the indent centre (slices 1 and 3 in Figure 9). The steep strain gradients immediately below the indent make the analysis of the rotation field in slice 2 difficult. The differences in rotation fields becomes clearer by comparing slices 1 and 3. Here it is seen that the rotation fields in the 1 dpa sample are slightly smaller in magnitude and more confined than in the unimplanted material. This is particularly noticeable for $\theta_x$ and $\theta_z$ and confirms the 3D confinement of deformation in the 1 dpa sample.

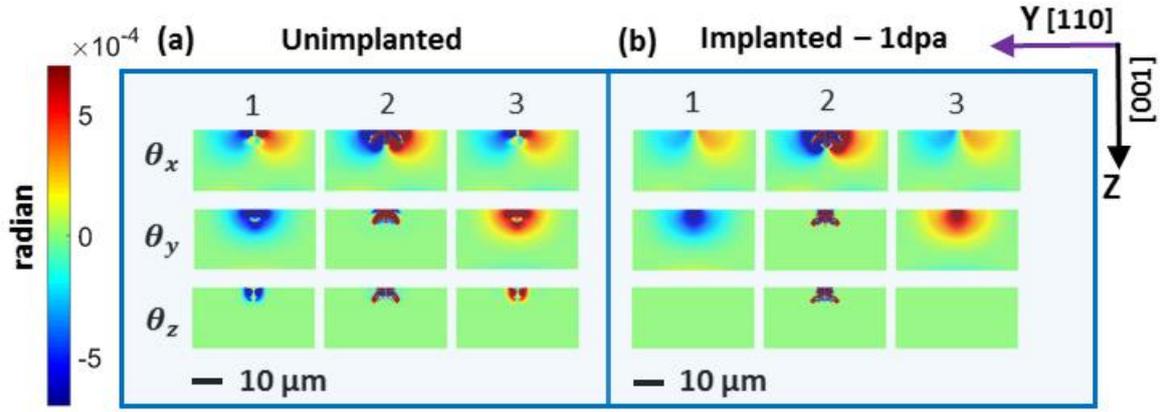

*Figure 9 – CPFE predictions of the lattice rotations underlying the indents for the (a) unimplanted sample and (b) the 1 dpa sample. The lattice rotations are plotted on virtual YZ slices where slice 2 is at the indent centre and slices 1 and 3 represents the YZ cross-sections 6 µm on either side of the indent centre. The lattice rotations measurements are shown in radian.*

The observations in Figure 8 and Figure 9 show that in the presence of irradiation damage, the applied deformation is accommodated in a more confined zone compared to the pristine sample. This implies that the effective strain gradients in the irradiated sample will be steeper than in the unirradiated case. The lattice curvature required to accommodate these strain gradients is provided by a distribution of geometrically necessary dislocations (GNDs) (Fleck and Hutchinson, 1997). Steeper strain gradients suggest a higher density of GNDs. It has been argued that GNDs play an important role in the nucleation of cracks (Chen et al., 2017; Stroh, 1957; Wan et al., 2014) and possibly in crack propagation through crack tip blunting (Jiang et al., 2016). Thus, considering that this dislocation-based stored energy could be useful criterion in determining material failure, we investigate the GND distribution in the irradiated sample and its variation with dose.

Details of the GND density computation can be found elsewhere (Suchandrima Das et al., 2018) and in Appendix F. Here we use the L2 minimisation technique, which minimizes the sum of squares of the dislocation densities ($\boldsymbol{\rho}$), i.e. $\sum_j \rho_j^2 = \boldsymbol{\rho}^T \boldsymbol{\rho}$, to compute the total GND density summed over the twelve a/2<111> {110} slip systems considered here (Marichal et al., 2013; Srivastava et al., 2013)). We assume that dislocations are either of pure edge or pure screw type, resulting in 16 dislocation types in total (Suchandrima Das et al., 2018).

The GND density produced across all slip systems as measured by HR-EBSD (Birosca et al., 2019; Chen et al., 2017; Zhu et al., 2017) and computed by CPFE is compared for the unimplanted, 0.01, 0.1, 0.32 and 1 dpa samples (Figure 10).

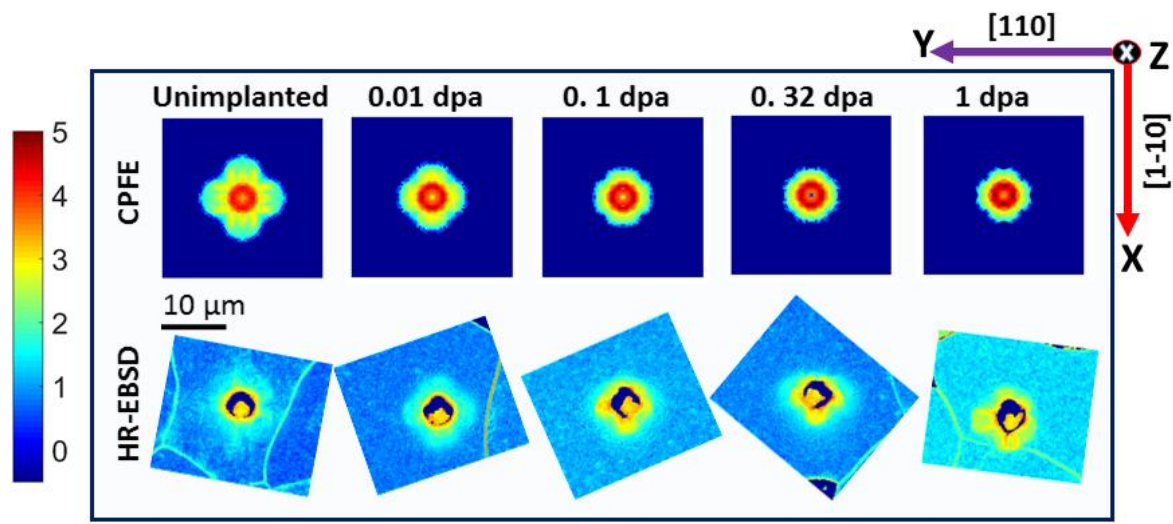

*Figure 10 – Sum of GNDs (ρ) over all slip systems as predicted by CPFE and measured by HR-EBSD for the self-ion implanted tungsten samples of the varying damage levels. The GNDs on each of the 12 slip systems were computed using L2 minimisation. The plots are shown on the sample surface i.e. on the XY cross-section. The same colour scale is used for all plots showing $\log_{10}(\rho)$ with ρ in $1/\mu m^2$. The HR-EBSD maps are rotated to have the same in-plane orientation as in the CPFE model.*

We note that the absolute values of GND densities calculated by the two techniques (HR-EBSD and CPFE) differ. This may be partly attributed to the instrumental broadening arising from the HR-EBSD measurement (details in Appendix F). Further, differences arise due to the different step size and mesh size used in HR-EBSD and CPFE respectively. Owing to the lack of inherent length scale in plasticity, the step size or mesh size is directly representative of the considered Burgers' circuit size. GNDs are the excess dislocations stored within the Burgers' circuit that are required to accommodate the lattice curvature (Ashby, 1970). With reduction in step size (or Burgers' circuit size) the fraction of dislocations that appear as GNDs increases (more dipoles are resolved as GNDs) until finally, for Burgers' circuit smaller than the dipole size, every dislocation appears as a GND. Thus the density of GNDs increases with decreasing step size, as shown by Jiang et al. in simulations of a deformed copper polycrystal (Jiang et al., 2015, 2013). The mesh size in CPFE was chosen to ensure there is no mesh dependence, with a smallest element size of 50 nm near the indenter tip. The step size in HR-EBSD (169 nm) was chosen to ensure that the resolution is sufficient to resolve strain gradients near indents, while reducing the measurement noise. The GND density data, predicted by CPFE, when determined using a step size equal to that used in HR-EBSD (169 nm) and blurred (to account for the instrumental broadening), shows a better quantitative agreement with the corresponding data from HR-EBSD. This is shown in Figure F.2 in Appendix F.

However, considering that GND density calculation is inherently affected by the step size or the chosen Burgers' circuit size, there is no absolute correct value that one may aim to achieve. Thus, quantitative agreement is not sought in Figure 10. Rather, Figure 10 demonstrates the agreement between HR-EBSD and CPFE, regarding the trend of dose-dependent change in GND density distribution. A clear confinement of the GND density field is evident with increasing damage dose up to 0.1 dpa. Beyond 0.1 dpa, there is little change in the GND density distribution, which is consistent with the mechanical response observed in Figure 5 and the observations of the indent surface profile Figure 6. The confinement of the GND density field beneath indents follows the same trend, as shown by the CPFE predicted GND density plots on the XZ and YZ cross-sections through the indent (Figure 1 (b) for reference co-ordinate frame) in Figure F.4 in Appendix F.

The orientation-dependence of GND density distribution is investigated through CPFE for the unimplanted and the 1 dpa sample. For these two cases, Figure 11 shows the GND density produced across all slip systems, as predicted by CPFE, for <001>, <011> and <111> oriented grains. For all grain orientations the deformation field becomes more confined near the indent in the implanted material, compared to a more widespread deformation zone in the unimplanted material. This result, together with Figure 7, confirms our hypothesis that the defect-dislocation interaction mechanism responsible for the modified behaviour of the implanted material i.e. the phenomenon of strain localisation, is orientation-independent.

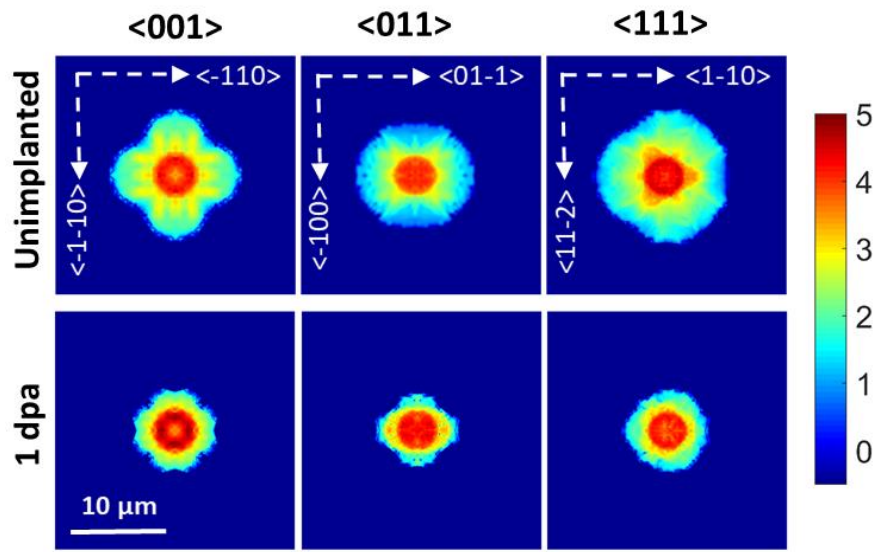

*Figure 11 - A comparison of sum of GNDs (ρ) over all slip systems as predicted by CPFE in all three grains for the unimplanted and the 1 dpa self-ion-implanted sample. GNDs on each of the 12 slip systems were computed using L2 minimisation. The plots are shown on the sample surface i.e. on the XY cross-section. The same colour scale is used for all plots showing $\log_{10}(\rho)$ with ρ in $1/\mu m^2$. The 10 μm scale bar applies to all plots. The in-plane orientations indicated for each orientation in the plots in the top row also apply to the corresponding plots in the bottom row.*

A notable difference between the HR-EBSD-measured and CPFE-predicted GND density distribution is the appearance of pronounced "lobes" around the indents at the higher damage levels (see HR-EBSD measurement of 1 dpa sample in Figure 10). This lobe formation may be due to the evolution of defect morphologies (such as clustering of defects or a long-range structural ordering of the defect microstructure) at higher damage levels, as captured in the experimental HR-EBSD measurements. While CPFE accurately captures the localisation of deformation at the higher dose levels, the absence of dose-dependent structural and/or organisational development of the defect microstructure and lack of integrated dislocation dynamics, may be rendering the model incapable of simulating the lobe formation as observed experimentally e.g. in the 1 dpa sample. Advanced CPFE models with integrated dislocation defect interactions are currently being developed by various groups. Examples of such models have been reported by Li et al. to simulate irradiation hardening in iron single crystal (Li et al., 2014) and by Ohashi et al. who simulated scale dependent characteristics of mechanical properties of polycrystals (Ohashi et al., 2007). However, to account for the increased complexity of the microstructure, an increase in the number of assumptions and fitting parameters is unavoidable.

The current simulations and experiments provide clear evidence for irradiation induced softening in self-ion implanted tungsten. The CPFE model captures the variation of mechanical properties across a range of damage levels, including damage saturation. Importantly, it does so with a minimal number of fitting parameters: $\tau_H^0$ is based on TEM observations, and $\gamma$ and $k$ are kept constant across all damage levels (fitted to the 0.01 dpa experimental result). Interestingly, the formulation closely follows the strain softening formulation adopted to simulate helium-implantation damage in tungsten (Das et al., 2019b). It is surprising that this relatively simple formulation is sufficiently general to allow the effects of vastly different damage microstructures (induced by helium and self-ion bombardment) to be simulated with good agreement with experimental results. There are a number of improvements that could be made, for e.g. accounting for the contribution of defects too small to be resolved by TEM. Also, insights drawn from atomistic and/or dislocation dynamics simulations that can give detailed information about the interaction between gliding dislocations and loops of varying sizes and morphologies (i.e. a more precise estimate of the obstacle strength and how its overcome) would further improve the accuracy of the model.

Despite these limitations, the current model provides an effective and straightforward means of exploring the impact of ion-induced damage on dislocation-mediated plastic slip as a function of dose. It triggers a promising idea, whereby minimal parameters derived from experimental observations from a small sample of the irradiated reactor component (e.g. TEM observations shedding light on neutron-irradiated loop concentration and estimates of the helium concentration) could be used in such predictive models to predict the anticipated macroscopic properties.

## 5.3. Translating small-scale response to predict macroscopic behaviour

Having verified the accuracy of the CPFE model for self-ion implanted tungsten, it can be used to predict the macroscopic deformation behaviour of a similarly irradiated, polycrystalline bulk material. To do this we generated a polycrystalline cube ($120 \times 120 \times 120$ μm$^3$) in Abaqus, with 512 cubic grains as shown in Figure 12 (a). Each grain was assigned a random crystallographic orientation and the elastic material properties of tungsten (see Appendix B). The bottom of the cube was kept fixed and symmetric boundary conditions were applied on the XY and YZ faces. The cube was meshed with C3D20R elements such that each grain was assigned 64 elements, giving a total mesh size of 32768 elements. Uniaxial compression tests were simulated by displacing the top surface in the negative Y-direction by 6 μm (5% strain applied). Crystal plasticity was integrated into the simulation by calling the previously developed UMAT at every Gauss point. The simulations were carried out for the unimplanted material, and 0.01, 0.1, 0.32 and 1 dpa damage levels, by using the relevant $\tau_H^0$ in the CPFE model in each case (Table 3 shows the dose-dependence of $\tau_H^0$. All other CPFE parameters are dose independent and are provided in Appendix C). $\sigma_{yy}$ and $U_{yy}$ (displacement in the Y direction) were extracted from all nodes of each element, for each time increment. At each time increment, the average compressive stress was considered to be $\sigma = \sum_n \sigma_{yy}/n$ where $n$ is the number of nodes. $U_{yy}$ and the original dimension of the cube (L = 120 μm) were used to compute the true strain at each time increment as $\varepsilon = \ln\left(\frac{L+U_{yy}}{L}\right)$.

An interesting question in these polycrystal simulations concerns the role of grain boundaries. Grain boundaries can act as strong obstacles to dislocation glide and thus fine-grained metals appear stronger (Essmann et al., 1968; Hirth and Lothe, 2016). The strain-gradient crystal plasticity formulation implemented here, captures this effect by accounting for the increased plastic strain-gradient due to mismatch of slip at the grain boundaries and thereby an increased population of GNDs contributing to hardening (Eq. (2)) (Fleck et al., 1994; Gao et al., 1999). Prior studies have also shown that grain boundaries act as efficient sinks for irradiation-induced defects (El-Atwani et al., 2017; Han et al., 2012; Wadsworth et al., 2002). However, the denuded zones being a few tens of nanometers thick, this only has a significant impact for nano-crystalline materials. Thus, this effect is not considered in the coarse-grained polycrystal simulations conducted here. Further detailed effects of the grain boundaries such as the formation of a grain boundary work-hardened layer or grain-size dependence of yield stress (Meyersm and Ashworth, 1982; Murr and Hecker, 1979) have not been included in the polycrystalline model simulated here, since they are not probed in the experiments.

In this section, our main aim is to qualitatively demonstrate the utility of experimentally validated CPFE formulations as a bridge between micro-mechanical response and expected macroscopic behaviour for samples where only small-scale experiments are feasible. In future these simulations could then be extended to include higher order effects, for example due to grain boundaries.

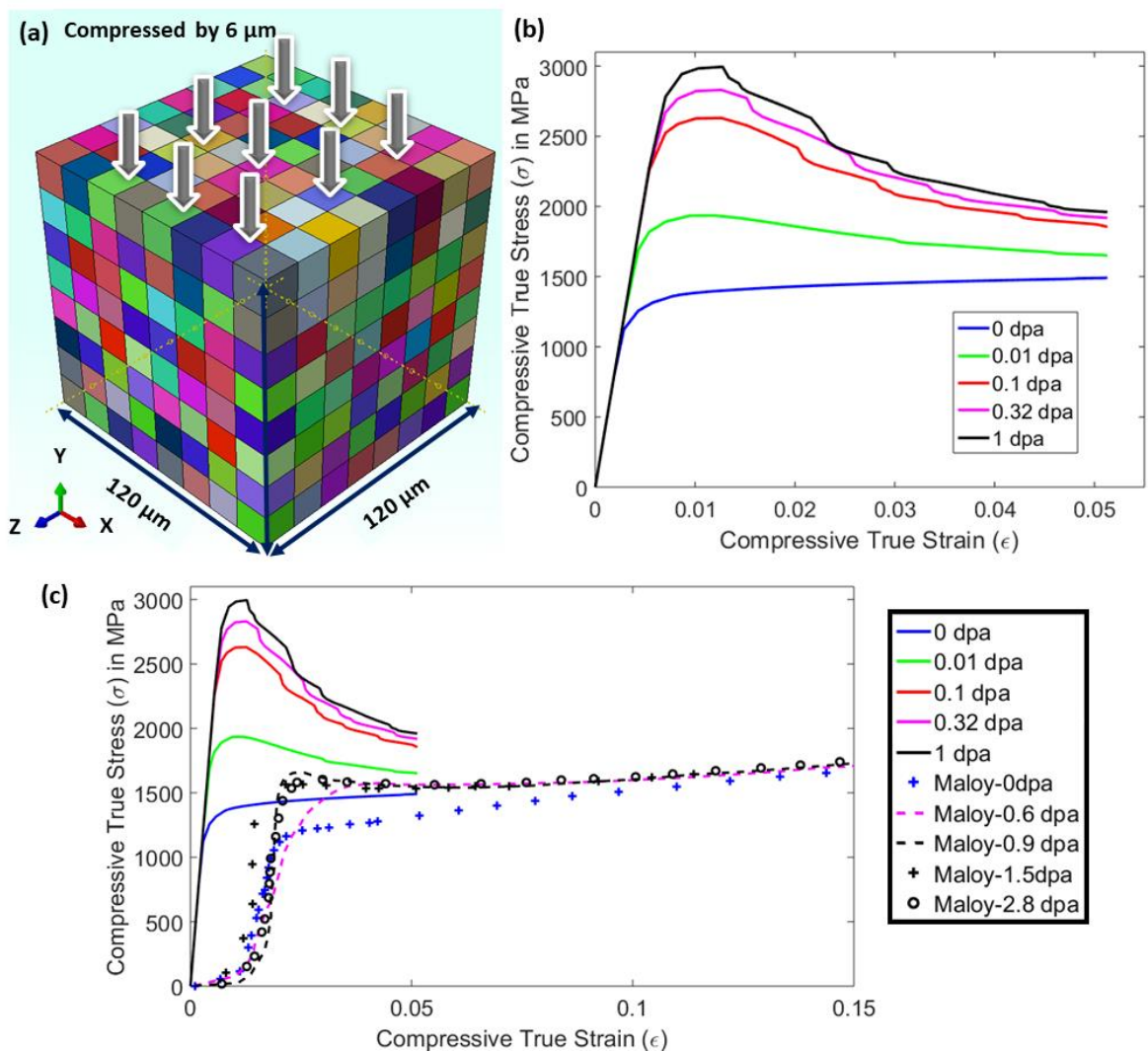

*Figure 12 – (a) Polycrystalline cube generated in Abaqus with 512 grains, each of which was assigned a random crystallographic orientation. (b) Macroscopic stress-strain curves generated by uniaxial compression testing of the polycrystalline cube. To generate each curve, the cube was assigned the crystal plasticity material model of self-ion implanted tungsten exposed to a particular damage level. (c) Curves from subplot (b) are shown here superimposed with data points acquired from Maloy et al. (Maloy et al., 2001). The data points from Maloy et al. correspond to compression tests on tungsten irradiated with protons at 60 °C (800 MeV, 1 mA proton beam) up to doses 0 dpa, 0.6 dpa, 0.9 dpa, 1.5 dpa and 2.8 dpa. The data points were taken from Figure 2 in the referred paper.*

The simulated compressive stress-strain curves predicted at each dose are shown in Figure 12 (b). They clearly show both irradiation-induced hardening and strain-softening increasing with dose and saturating beyond 0.1 dpa. This is in qualitative agreement with the compressive stress-strain curves obtained for proton-irradiated tungsten (800 MeV, 1 mA proton beam, 60°C) (Maloy et al., 2001) as shown in Figure 12 (c) (data points have been extracted from Figure 2 in the referred paper). Figure 12 (c) shows that, akin to predictions made by our CPFE model, in the data obtained by Maloy et al. there is irradiation-induced hardening that saturates beyond a certain dose (0.9 dpa in this case). The ultimate convergence of the stress-strain curves of proton-irradiated tungsten with that of pure tungsten, observed experimentally by Maloy et al. (Figure 12 (c)), qualitatively validates the irradiation-induced strain-softening as seen in the stress-strain curves predicted by CPFE for the self-ion implanted tungsten samples.

With such established crystal plasticity material model, further tests, such as cyclic loading of such polycrystalline irradiated materials, may be simulated to assess their fatigue life (Korsunsky et al., 2007) or thermo-mechanical analysis at fusion relevant steady/transient heating loads (Fukuda et al., 2015; Li et al., 2017) may be performed. Such predictions, made as a function of evolving dose, could then be integrated into the reactor design to ensure property evolution is correctly accounted for when assessing long-term serviceability of components.

## 6. Conclusion

Understanding the impact of in-service radiation damage on structural and functional material properties is vital for the design of future fusion reactors. Estimating macroscopic properties is challenging as ion-implanted samples, used to mimic neutron irradiation, are only a few microns thick. Here we have demonstrated how experimental characterisation data from ion-implanted samples can be used to predict the anticipated macroscopic mechanical properties using a two-step process: First we develop a mesoscale material model that captures the physics of the irradiation damage. The numerical formulation for the material model is based on and validated against a combination of experimental observations on ion-implanted samples. This material model is then used to simulate the macroscopic deformation behaviour of similarly irradiated polycrystalline bulk material.

We have used self-ion implanted tungsten as a prototypical material to demonstrate this process. The results allow the following conclusions to be drawn about the dose-dependent deformation behaviour of self-ion-implanted tungsten:

- Surface profiles of indents in self-ion implanted samples may show pile-up or suppression of pile-up as a function of grain orientation. In self-ion-implanted tungsten the largest pile-up occurred in <001> oriented grains, similar to observations made in helium-implanted tungsten.

- Pile-up around indents in <001> grains increased with increasing damage level and reached a saturation beyond 0.032 dpa.

- A similar saturation is observed in the mechanical response, captured by nano-indentation load-displacement curves. The maximum load increases by ~27% for damage levels between 0.032 and 1 dpa.

- Irradiation-induced hardening accompanied by surface pile-up and slip step formation, indicates that irradiation defects initially act as strong obstacles to glide dislocations, but that their obstacle strength is reduced with increasing plastic deformation. A CPFE formulation built on this hypothesis correctly predicts indentation load and indent surface profiles for unimplanted, 0.01, 0.1, 0.32 and 1 dpa samples. The shear resistance of irradiation defects in CPFE is physically-based and derived from TEM observations. Only two parameters are fitted to the experimental results of the 0.01 dpa sample and kept constant for all other damage levels.

- CPFE correctly captures the dependence of pile-up on the grain orientation as exemplified through the case of the unimplanted and 1 dpa implanted samples. Our results show that the underlying interaction mechanism between irradiation defect and glide dislocation is orientation independent. Differences in pile-up morphology are simply due to the relative orientation of the crystal lattice with respect to the sample surface and the spherical indenter tip.

- Confinement of the deformation field is explored in detail by comparing the unimplanted and the 1 dpa sample. CPFE predictions of residual lattice rotations and lattice strains around the indents are in good agreement with HR-EBSD measurements.

- CPFE and HR-EBSD computed GND density fields around the indents, across a range of damage levels, show similar trends: deformation becomes increasingly confined with rising damage level up to 0.1 dpa. Beyond this little change is noticed.

- The CPFE model was successfully used to predict the macroscopic deformation behaviour of self-ion irradiated polycrystalline tungsten. The macroscopic stress-strain curves show initial irradiation-induced hardening, followed by strain-softening during deformation. Similar to the small-scale behaviour, irradiation induced changes were seen to increase as a function of dose and saturate beyond 0.1 dpa.

## Acknowledgements

We thank A.J. Wilkinson for providing the software for HR-EBSD analysis. This work was funded by Leverhulme Trust Research Project Grant RPG-2016-190. ET acknowledges financial support from the Engineering and Physical Sciences Research Council Fellowship grant EP/N007239/1. Electron and atomic force microscopy were performed at the David Cockayne Centre for Electron Microscopy, Department of Materials, and at the LIMA lab, Department of Engineering Science, both at the University of Oxford. Ion implantations were performed at the Helsinki Accelerator Laboratory, Department of Physics, University of Helsinki.

# Appendix A

*Table A.1 - List of the out-of-plane orientation of the chosen point in each sample grain and the misorientation of the chosen point with respect to the perfect <111> ,<110> or <001> out-of-plane direction.*

| Sample (grain is <001> unless mentioned otherwise) | Euler Angles $(\varphi_1, \varphi, \varphi_2)$[3] | Out-of-plane orientation | Misorientation with <111> ,<110> or <001> (in degrees) |
|---|---|---|---|
| Unimplanted grain 1 | 277.8,5.4,83 | [9.34,1.15,99.56] | 5.4 |
| Unimplanted grain 2 | 71.8,10.5,23.4 | [7.24,16.72,98.33] | 10.5 |
| Unimplanted grain 3 | 180.9,6.6,29.6 | [5.68,9.99,99.34] | 6.6 |
| Unimplanted grain 4 <011> | 254.5,39.7,6.8 | [7.56,63.43,76.94] | 6.99 |
| Unimplanted grain 5 <111> | 330.3,43.5,61.6 | [60.55,32.74,72.54] | 16.78 |
| 0.01 dpa grain 1 | 214.7,1.6,293.3 | [-2.56,1.1,99.96] | 1.6 |
| 0.01 dpa grain 2 | 33.7,2.4,176.4 | [0.26,-4.18,99.91] | 2.4 |
| 0.01 dpa grain 3 | 51.2,9.1,159 | [5.67,-14.77,98.74] | 9.1 |
| 0.018 dpa grain 1 | 190.3,5.4,21 | [3.37,8.79,99.56] | 5.4 |
| 0.018 dpa grain 2 | 289,7.8,69 | [12.67,4.86,99.07] | 7.8 |
| 0.018 dpa grain 3 | 213.4,11.2,15.8 | [5.29,18.69,98.1] | 11.2 |
| 0.032 dpa grain 1 | 268.3,9.6,92 | [16.67,-0.58,98.6] | 9.6 |
| 0.032 dpa grain 2 | 185.8,3.2,19 | [1.82,5.28,99.84] | 3.2 |
| 0.032 dpa grain 3 | 13.4,5.8,198.2 | [-3.16,-9.6,99.4] | 5.8 |
| 0.1 dpa grain 1 | 27.8,3.2,181.1 | [-0.11,-5.58,99.8] | 3.2 |

[3] The Euler angle convention used is as follows: $Z1 = \begin{bmatrix} \cos\varphi_1 & \sin\varphi_1 & 0 \\ -\sin\varphi_1 & \cos\varphi_1 & 0 \\ 0 & 0 & 1 \end{bmatrix}$; $X = \begin{bmatrix} 1 & 0 & 0 \\ 0 & \cos\varphi & \sin\varphi \\ 0 & -\sin\varphi & \cos\varphi \end{bmatrix}$; $Z2 = \begin{bmatrix} \cos\varphi_2 & \sin\varphi_2 & 0 \\ -\sin\varphi_2 & \cos\varphi_2 & 0 \\ 0 & 0 & 1 \end{bmatrix}$ and the rotation matrix R = Z1 * X * Z2.

| | | | |
|---|---|---|---|
| 0.1 dpa grain 2 | 16.3,7.4,200 | [-4.41,-12.1,99.1] | 7.4 |
| 0.1 dpa grain 3 | 162.1,4.1,341.8 | [-2.23,6.79,99.74] | 4.1 |
| 0.32 dpa grain 1 | 103.8,4.9,64.4 | [7.7,3.69,99.63] | 4.9 |
| 0.32 dpa grain 2 | 59.6,9.2,114.6 | [14.54,-6.66,98.71] | 9.2 |
| 0.32 dpa grain 3 | 344.1,10.8,230.3 | [-14.42,-11.97,98.2] | 10.8 |
| 1 dpa grain 1 | 126.5,7.6,67.4 | [12.21,5.08,99.12] | 7.6 |
| 1 dpa grain 2 | 319.8,7.1,87.9 | [12.35,0.45,99.23] | 7.1 |
| 1 dpa grain 3 | 207.5,5.1,0.4 | [0.06,8.89,99.6] | 5.1 |
| 1 dpa grain 4 <011> | 263.7,42.8,82.4 | [67.35,8.99,73.37] | 5.7075 |
| 1 dpa grain 5 <111> | 264.2,42.8,82.4 | [58.07,37.43,72.3] | 14.36 |

# Appendix B

The stiffness of tungsten (material for sample block) is ~36% that of diamond (material of the indenter tip) (Table B.1). It has been seen from simulations using a rigid sharp indenter, that simulation results match the experimental results well if the results are scaled using an effective modulus, $E_{eff}$ (M. Li et al., 2009)

$$P\,(exp.) = \frac{E_{eff\,(exp.)}}{E\,(FEA)}\,P(FEA) \qquad (B.1)$$

Thus, to avoid a full meshing and increase in simulation size, the indenter was designed as a discrete rigid wire frame.

*Table B.1 – Values of Young's modulus and Poisson's ratio for diamond (indenter tip) and tungsten (indented sample) as obtained from literature* (Bolef and De Klerk, 1962; Featherston and Neighbours, 1963; Klein and Cardinale, 1993)[4].

| $E_{diamond}$ | $E_{tungsten}$ | $\nu_{diamond}$ | $\nu_{tungsten}$ | $E_{eff}$ | $R_{indenter}$ |
|---|---|---|---|---|---|
| 1143 GPa | 410 GPa | 0.0691 | 0.28 | 322.58 GPa | 4.2 μm |

---

[4] With the assumption of isotropic, linear elastic solid, the Young's modulus and Poisson's ratio are related to the elastic constant as follows: $E = c_{11} - 2\left(\frac{c_{12}^2}{c_{11}+c_{12}}\right)$ and $\nu = c_{12}/(c_{11}+c_{12})$.

# Appendix C

*Table C.1– List of parameters used in the constitutive law in the CPFE formulation and their corresponding values.*

| Material Property | Value | Reference |
|---|---|---|
| Elastic modulus $E$ | 410 GPa | (Ayres et al., 1975; Bolef and De Klerk, 1962; Featherston and Neighbours, 1963; Klein and Cardinale, 1993) |
| Shear modulus $G$ | 164.4 GPa | (Ayres et al., 1975; Bolef and De Klerk, 1962; Featherston and Neighbours, 1963; Klein and Cardinale, 1993) |
| Poisson's ratio $v$ | 0.28 | (Ayres et al., 1975; Bolef and De Klerk, 1962; Featherston and Neighbours, 1963; Klein and Cardinale, 1993) |
| Burgers' vector $b$ | $2.7 \times 10^{-10}$ m | (Dutta and Dayal, 1963) |
| Activation energy $\Delta F$ | 0.22 eV | Section C.1 |
| Boltzmann constant $k$ | $1.381 \times 10^{-23}$ J/K | (Sweeney et al., 2013) |
| Temperature $T$ | 298 K | Room temperature assumed similar to experimental conditions |
| Attempt frequency $v$ | $1 \times 10^{11}$ s$^{-1}$ | Value chosen and kept fixed |
| Density of statistically stored dislocations, $\rho_{SSD}$ | $1 \times 10^{10}$ m$^{-2}$ | Section C.2 |
| Density of mobile dislocations $\rho_m$ | $1 \times 10^{10}$ m$^{-2}$ | Section C.2 |
| Probability of pinning $\Psi$ | $0.857 \times 10^{-2}$ | Value chosen and kept fixed |
| $\tau_c^0$ | 500 MPa | Fitted to experimental data of unimplanted sample |
| $\gamma$ | 0.08 | Fitted to experimental data of 0.01 dpa sample and kept fixed for all other dose rates |
| $C'$ | 0.1 | Fitted to experimental data of unimplanted sample |
| Average barrier strength of the implantation-induced loops $m$ | 1.23 | Fitted to experimental data of 0.01 dpa sample and kept fixed for all other dose rates |
| $\tau_H^0$ (0.01 dpa) | 260 | Derived based on TEM data |
| $\tau_H^0$ (0.1 dpa) | 588 | Derived based on TEM data |
| $\tau_H^0$ (0.32 dpa) | 679 | Derived based on TEM data |
| $\tau_H^0$ (1 dpa) | 766 | Derived based on TEM data |

### Section C.1:

Nano-indentation up to 500 nm depth was carried out on unimplanted tungsten grains of <001> orientation at three different rates as shown in Figure C.1(a), where rate 3 corresponds to the rate used for all other nano-indentation tests reported in this study (Figure 5). Superimposition of the load-displacement curves obtained at the three different strain rates show that the deformation behaviour in tungsten is insensitive to strain rate at these low deformation rates (Figure C.1 (b)).

To incorporate this strain-rate insensitivity in the CPFE formulation we focussed on the activation energy ($\Delta F$). It has been seen that at these low strain rates ($\dot{\varepsilon} < 10^3\,\text{s}^{-1}$), the primary factor responsible for controlling strain-rate sensitivity is the activation energy or the rate of thermally activated processes allowing the escape of dislocations pinned at obstacles (Zheng et al., 2016). The range of activation energy considered feasible for tungsten varies in the order of $10^{-19}$ J and $10^{-21}$ J (Cottrell, 1990). A lower activation energy implies a lower energy barrier required for the escape of the pinned dislocation (Gibbs, 1969) and thus a higher probability of being unaffected by varying strain rates (Zheng et al., 2016). This together with the empirical evidence of activation energy of ~ 0.2 eV (Gumbsch, 1998) in tungsten, led us to value ($\Delta F$ at $3.4559 \times 10^{-20}$ J or ~0.22 eV in the CPFE formulation.

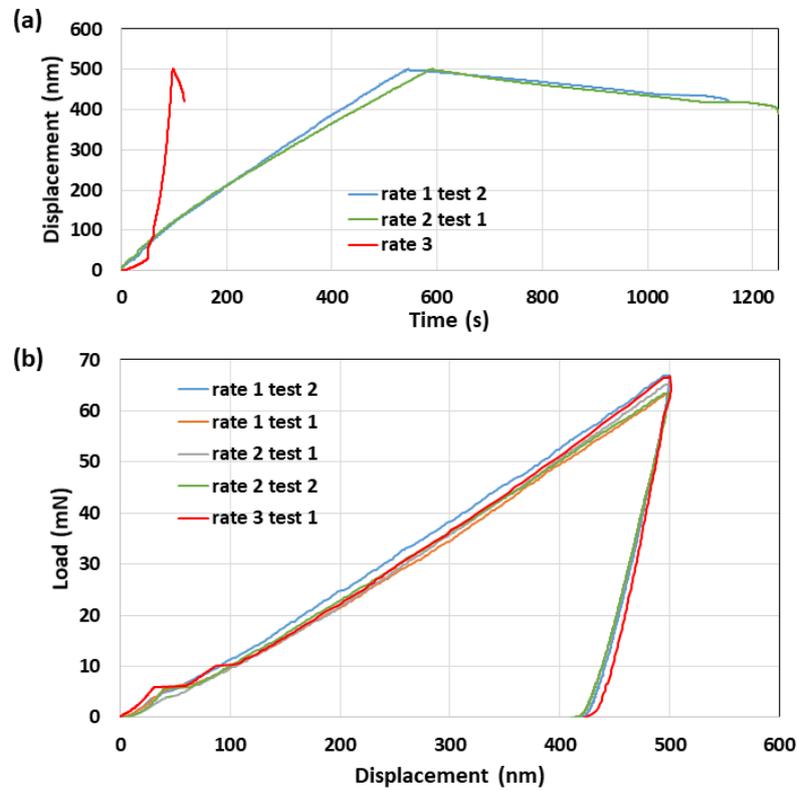

*Figure C.1 – (a) Plot showing three different rates at which nano-indentation was performed on unimplanted tungsten grains of <001> orientation. (b) Plots of the load-displacement curves corresponding to the applied strain rates in (a).*

## Section C.2: Determining values of $\rho_{SSD}$ and $\rho_m$

In nano-indented pure tungsten, the GND density was found to be on the order of ~ $10^{17}$ m$^{-2}$ by both Laue diffraction measurements and CPFE simulations (Suchandrima Das et al., 2018). With reference to this, for the present CPFE calculations, we assume the SSD densities, $\rho_{SSD}$ and $\rho_m$, to be much smaller than the GND density estimate, on the order of $10^{10}$ m$^{-2}$.

# Appendix D

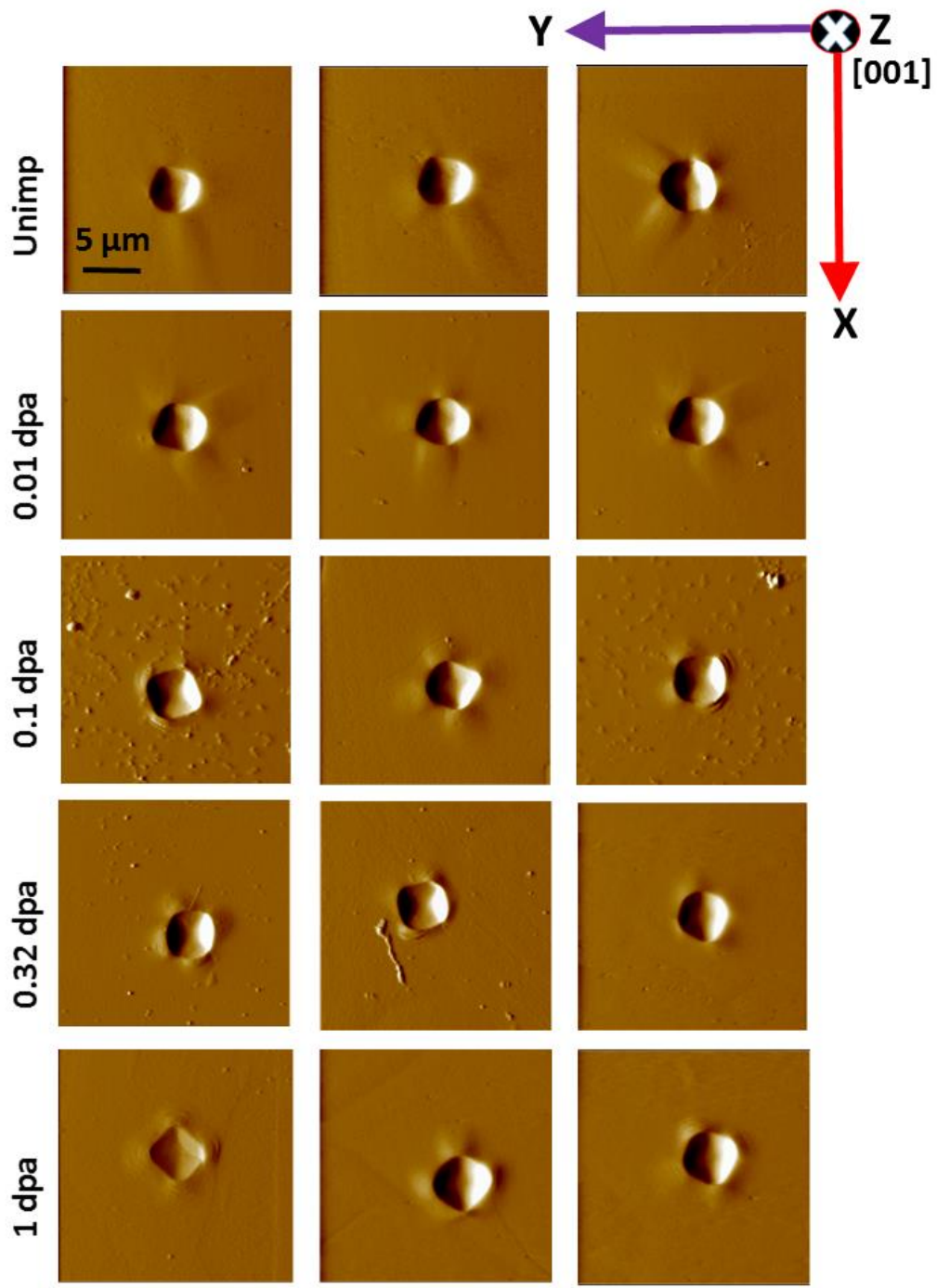

*Figure D.1 – AFM gradient micrographs of indents in three <001> grains in self-ion implanted tungsten samples of varying damage levels.*

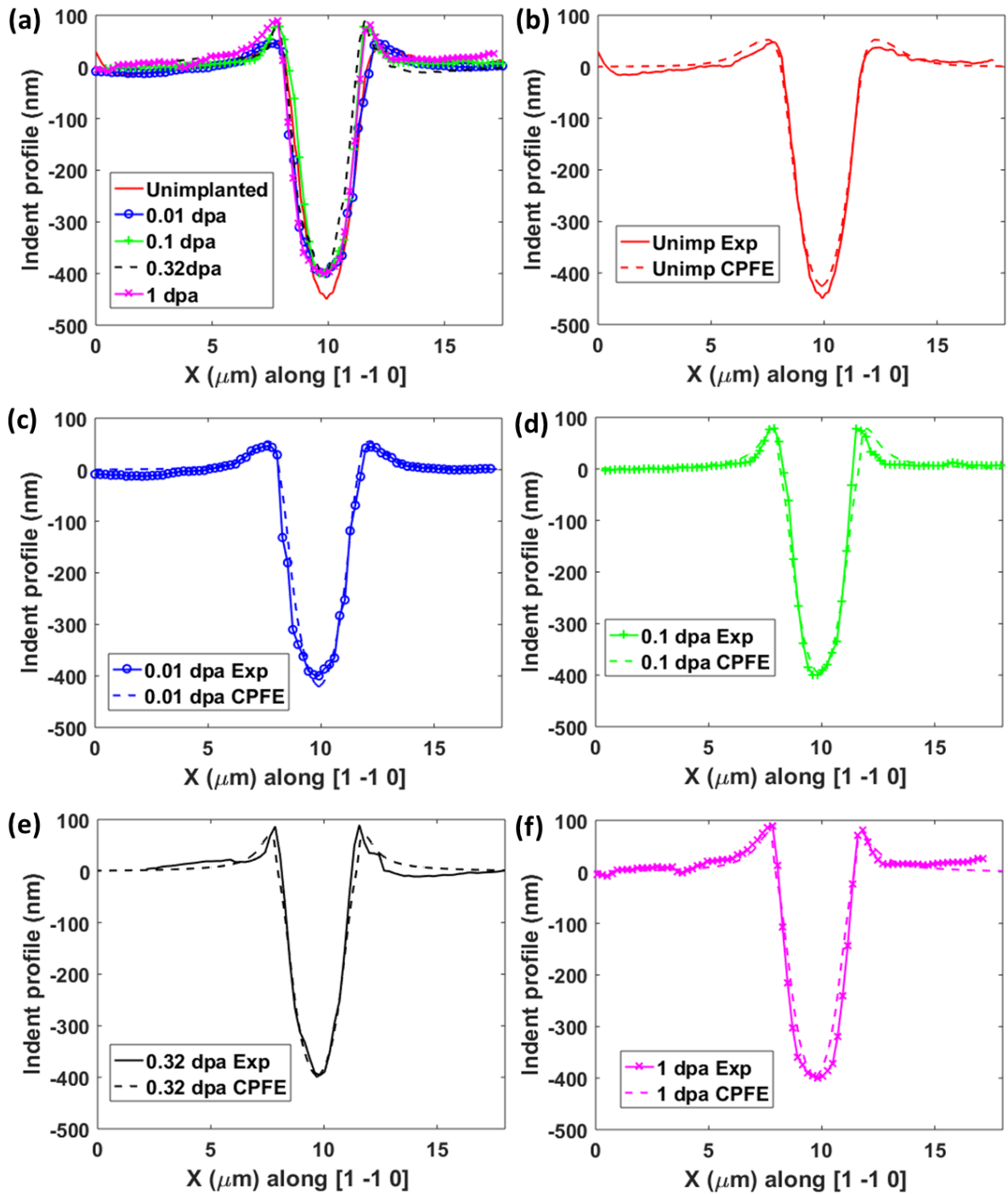

*Figure D.2 – (a) Line plots drawn through the <110> directions of the surface profile of the indent as measured by AFM (Figure 6 top row) for self-ion-implanted samples of doses between 0 and 1 dpa. (b) – (f) Superimposition of line plots drawn through <110> directions of the surface profile measured by AFM (Figure 6 top row) and predicted by CPFE (Figure 6 bottom row) for (b) pure tungsten, (c) 0.01 dpa, (d) 0.1 dpa, (e) 0.32 dpa and (f) 1 dpa.*

# Appendix E

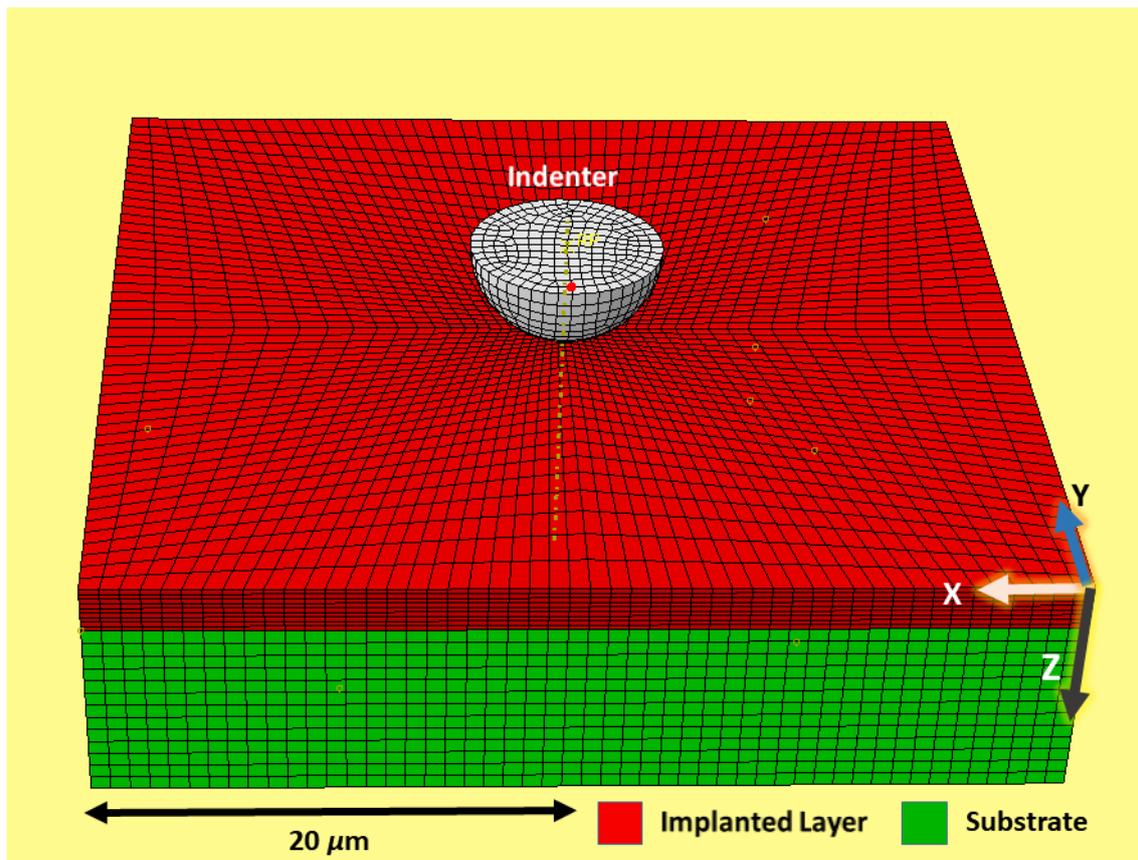

*Figure E.1 - CPFE mesh of the full model created in Abaqus for simulation of the nano-indentation experiments for the <111> oriented grains.*

# Appendix F

## Section F.1: Comparison of lattice distortions around indents in 1 dpa and 0.32 dpa sample using HR-EBSD

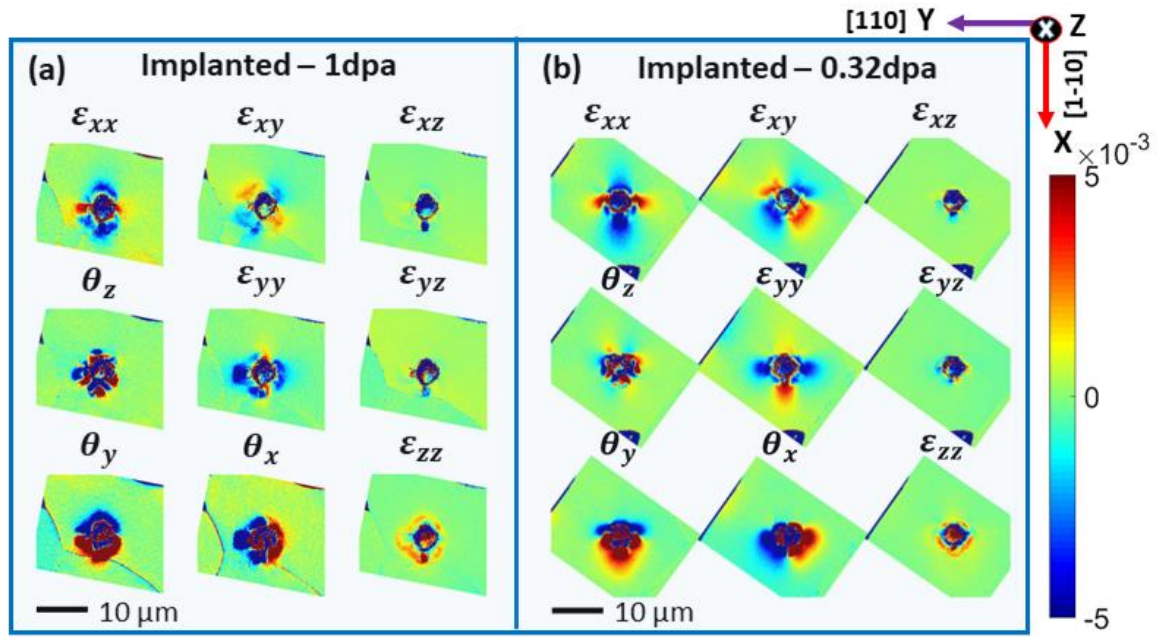

*Figure F.1 - HR-EBSD measurement of lattice rotations and all six components of the residual elastic deviatoric lattice strain plotted on the XY plane (indent surface) for the (b) 1 dpa sample and (b) 0.32 dpa sample.*

## Section F.2: Details of GND density calculation

GND density provides a useful measure of the combined effect of the lattice rotation and lattice strain. It is a direct function of the total plastic deformation gradient, $F^p$ (or elastic deformation gradient, $F^e$) (Suchandrima Das et al., 2018). Details of the GND density computation can be found elsewhere (Suchandrima Das et al., 2018). Briefly, GND density, $\rho$ is calculated by exploiting its relation with the closure failure ($<B>$) induced by plastic deformation in a material, where $<B> = \iint_S (\text{CURL}\,(F^p))^T\,NdS \cong \iint_S (-\text{CURL}\,(F^e))^T\,NdS$ (Nye, 1953). The relation can be defined as

$$\sum_\lambda (\boldsymbol{b}^\lambda \otimes \boldsymbol{\rho}^\lambda) = (CURL\ (\boldsymbol{F}^p))^T = (-CURL\ (\boldsymbol{F}^e))^T \quad (E.1)$$
$$\cong (-CURL(\boldsymbol{\varepsilon}^e + \boldsymbol{\omega}^e))^T$$

where, $\boldsymbol{b}$ is the Burgers' vector, $\lambda$ is one of the $n$ slip systems, CURL of any second-order tensor $V$ is described by $(\nabla \times \boldsymbol{V})_{km} = \epsilon_{ijk}\ V_{mj,i}$ (Suchandrima Das et al., 2018), $\boldsymbol{\varepsilon}^e$ is the residual elastic lattice strain and $\boldsymbol{\omega}^e$ is the residual lattice rotation. The 3 × 3 tensor obtained by calculating $(-CURL(\boldsymbol{\varepsilon}^e + \boldsymbol{\omega}^e))^T$, can be reshaped into a 9 × 1 column vector. $\boldsymbol{\rho}$ can be represented as a column vector representing densities of $j$ dislocation types. Then Eq. E.1 can be re-written as

$$\boldsymbol{A\rho} = \boldsymbol{\alpha} \quad (E.2)$$

where $\boldsymbol{A}$ is a linear operator (9×$j$ matrix, for $j$ types of dislocations), where the $j$[th] column contains the dyadic product of the Burgers' vector and line direction of the $j$[th] dislocation type. Usually $j > 9$, and thus there is no unique solution for $\boldsymbol{\rho}$. Hence, an optimisation technique needs to be used to obtain $\boldsymbol{\rho}$. In a recent study it was seen that L2 minimisation, which minimises the sum of squares of dislocation densities i.e. $\sum_j \rho_j^2 = \boldsymbol{\rho}^T.\boldsymbol{\rho}$, can reliably predict the sum of GNDs produced across all slip systems (Suchandrima Das et al., 2018). Here we use this minimisation technique as we are concerned primarily with the sum of dislocation densities produced across the considered slip systems (a/2<111> Burgers' vector gliding on {110} planes (Marichal et al., 2013; Srivastava et al., 2013)).

### Section F.3: Differences in GND density between HR-EBSD measurements and CPFE predictions

Figure F.1 (a) shows the GND density as measured by HR-EBSD for pure tungsten using a step size of 169 nm (same as shown in Figure 10). Figure F.1 (b) shows the GND density data predicted by CPFE for pure tungsten, using the defined mesh, with the smallest element size of 50 nm (same as shown in Figure 10). The observed quantitative disagreement between Figure F.1 (a) and (b) could arise from the difference in the chosen Burgers' circuit size for CPFE (50 nm) and HR-EBSD (169 nm). To investigate this, the GND density from CPFE (Figure F.1 (b)) was determined using a step size equal to that used in HR-EBSD, as shown in Figure F.1 (c). Comparison of Figure F.1 (a) and (c) still shows some quantitative differences which may be due to instrumental broadening from EBSD measurements (Rice et al., 2017). To investigate this, image blurring was applied to the GND density map (Figure F.1 (c)). Different blurring approaches were attempted using MATLAB; using a disk filter of different radii (using the 'imfilter' function) and using a 2-D Gaussian smoothing kernel with varying standard deviation (using the 'imgaussfilt' function). Blurring using a disk filter of radius 4 or blurring with a Gaussian smoothing kernel with 0.5 standard deviation produced similar results (Figure F.1 (d)) and showed the closest match to the GND density measured from HR-EBSD. Based on these investigations, we conclude that the differences in GND density computation between HR-EBSD measurements and CPFE predictions, as observed in Figure 10, may arise as a result of two factors:

1. Difference in chosen Burgers' circuit size.
2. Instrumental broadening

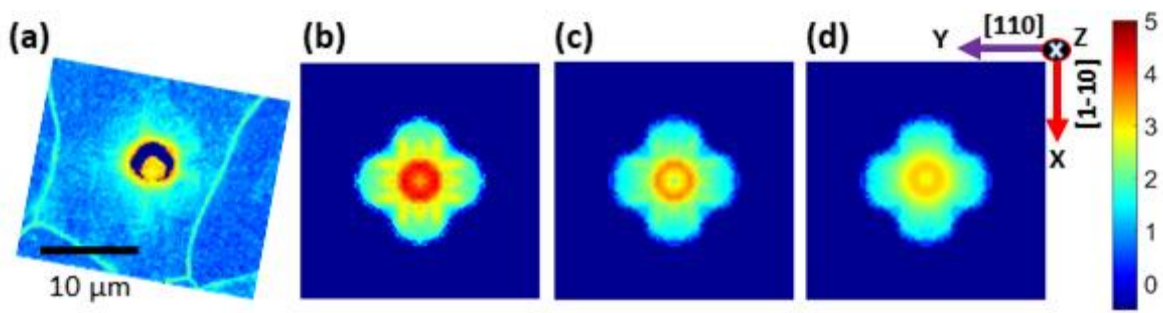

*Figure F.2 - Sum of GNDs (ρ) over all slip systems for unimplanted tungsten. (a) As measured by HR-EBSD (b) GND density predicted by CPFE mesh with smallest element size of 50 nm (as in Figure 10, unimplanted) (c) predicted GND density using a step size of 169 nm and (d) figure (c) after application of image blurring. The GNDs on each of the 12 slip systems were computed using L2 minimisation. The plots are shown on the sample surface i.e. on the XY cross-section. The same colour scale is used for all plots showing $\log_{10}(ρ)$ with ρ in $1/μm^2$. The HR-EBSD maps are rotated to have the same in-plane orientation as in the CPFE model.*

The impact of a similar blurring or instrumental broadening effect was also investigated for the comparative analysis of lattice distortions between experiments and simulations, by using the case of the 1 dpa sample. Figure F.3 (a) shows the lattice rotations and strains measured by HR-EBSD. Figure F.3 (b) and (c) show the corresponding predictions from CPFE without and with blurring effect respectively. Comparison of Figure F.3 (b) and (c) reveals that the applied blurring has little effect.

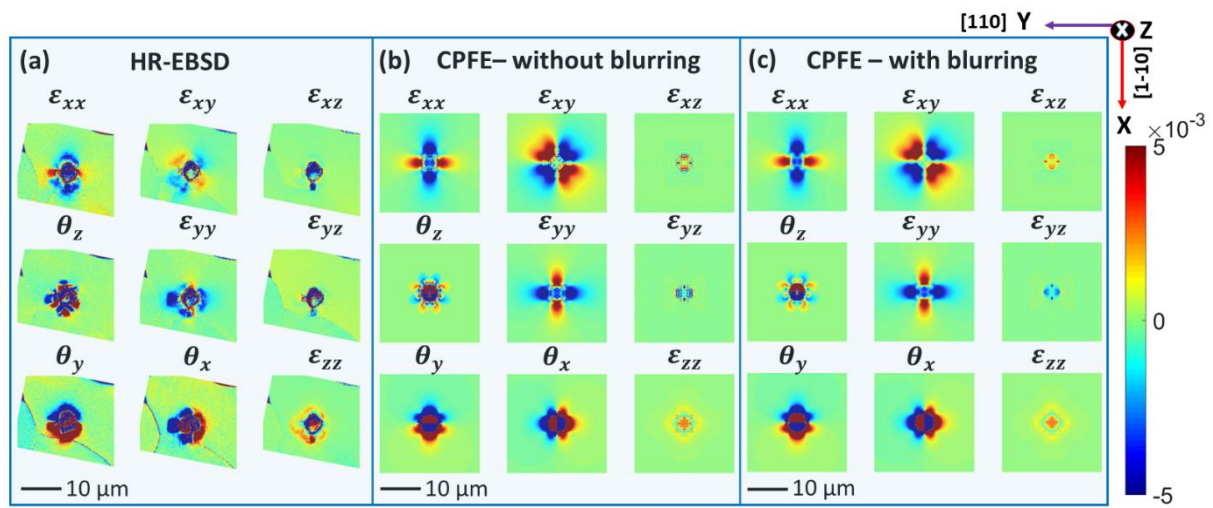

*Figure F.3 - HR-EBSD measurement of lattice rotations and all six components of the residual elastic deviatoric lattice strain plotted on the XY plane (indent surface) for the (a) 1 dpa sample. CPFE predictions of lattice rotations and all six components of the residual elastic deviatoric lattice strain plotted on the XY plane (sample surface) for the (b) 1 dpa sample. (c) Subplot (b) after being subjected to blurring effect.*

This explains the following:

a) The phenomenon of instrumental broadening though applicable to lattice distortions measured by HR-EBSD, has little effect when considered alone. As such, even without considering this effect, good agreement between experiments and simulations is observed for the lattice distortions (Figure F.3).

b) Marked differences between computations from experiments and simulations (as seen for GND density) occur only when the phenomenon of instrumental broadening is combined with change of step size (a phenomenon only specifically considered for the GND density, as it is directly involved in the computation of the same).

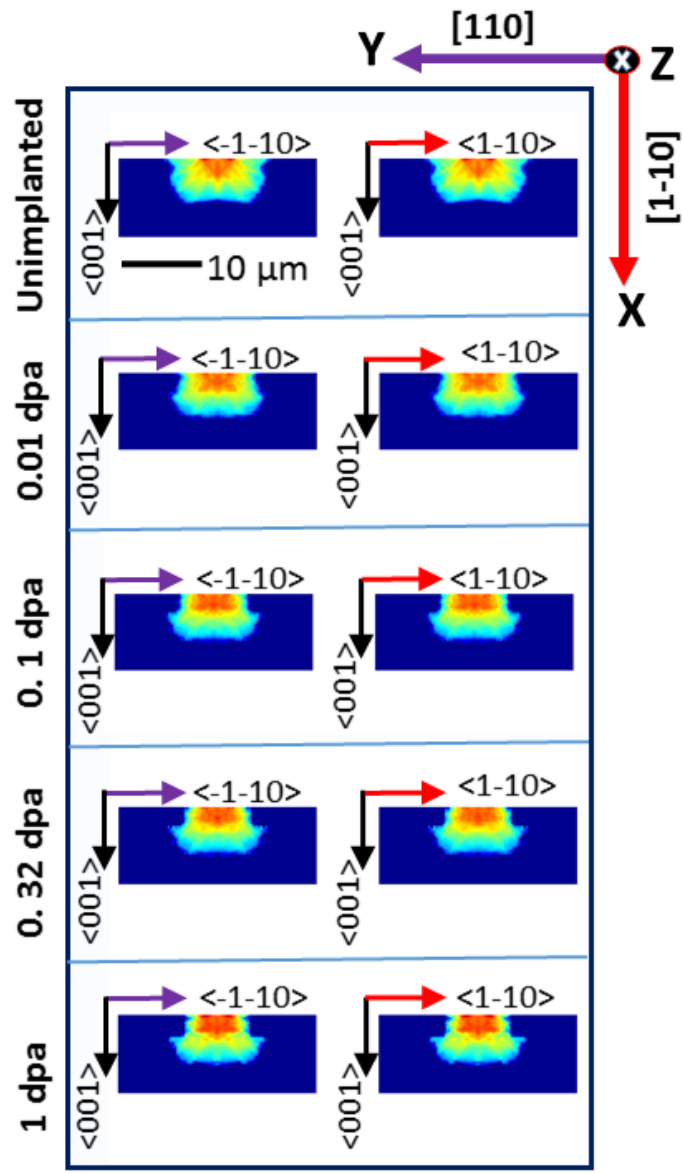

*Figure F.4 - Sum of GNDs (ρ) over all slip systems as predicted by CPFE for the self-ion implanted tungsten samples of the varying damage levels. The GNDs on each of the 12 slip systems were computed by L2 minimisation. The XZ and YZ cross-sections of 3D field of GND sum around the indent, as predicted by CPFE, are shown for each sample. The same colour scale is used for all plots showing $log_{10}(ρ)$ with ρ in $1/\mu m^2$. The XYZ coordinate system on the top right corner refers to the co-ordinate system used for the model and the experimental sample when looking at the sample surface from above.*